\documentclass[preprint,showpacs,preprintnumbers,amsmath,amssymb]{revtex4}
\usepackage{graphics,epsfig,subfigure}
\usepackage{color}
\usepackage{multirow}
\usepackage{array}
\usepackage{CJK}
\usepackage{ulem}
\usepackage{indentfirst}

\newcommand\beq{\begin{equation}}
\newcommand\eeq{\end{equation}}
\newcommand\beqn{\begin{eqnarray}}
\newcommand\eeqn{\end{eqnarray}}

\begin{document}

\title{Fermion localization mechanism with derivative geometrical coupling on branes}

\author{Yan-Yan Li\footnote{liyy2015@lzu.edu.cn}}
\author{Yu-Peng Zhang\footnote{zhangyupeng14@lzu.edu.cn}}
\author{Wen-Di Guo\footnote{guowd14@lzu.edu.cn}}
\author{Yu-Xiao Liu\footnote{liuyx@lzu.edu.cn, corresponding author}}
\affiliation{Institute of Theoretical Physics,
              ~Lanzhou University, Lanzhou 730000, P. R. China}

\begin{abstract}
  In order to localize fermions on branes with codimension one, one usually introduces the Yukawa coupling between fermions and background scalar fields or the recently proposed derivative fermion-scalar coupling in [Phys. Rev. D 89 (2014) 086001]. In this paper, we explore the coupling between a spinor field $\Psi$ and the scalar curvature of spacetime $R$, $\eta\Psi\Gamma^M\partial_M F(R)\gamma^5\bar{\Psi}$ with $F(R)$ a function of $R$, to investigate localization of the fermion. Because of
  $Z_2$ symmetry of the extra dimension, the new coupling mechanism proposed here can easily deal with the problem encountered in the Yukawa coupling with even background scalar fields.
  More importantly, the new mechanism will also work for the branes without background scalar fields.
  By investigating three examples, we find that fermions can be localized on the branes with the new mechanism.
\end{abstract}






\maketitle

\section{Introduction}
In the last two decades, the idea that our universe is embedded in a higher dimensional spacetime has attracted much  attention since Arkani-Hamed, Dimopoulos and Dvali (ADD) \cite{Arkani-Hamed1998} and Randall-Sundrum (RS) models \cite{Randall1999,Randall1999b} were presented.
Extra dimensions not only provide a new view of spacetime, but also can solve some outstanding problems that the standard model cannot deal with, such as dark matter \cite{Okada2004,Nihei2005}, dark energy \cite{Sahni2003,Arkani-Hamed1999}, gauge hierarchy problem \cite{Randall1999,Guo2015b,Yang2012,Antoniadis1998,Arkani-Hamed1998,Gogberashvili2002}, and cosmology constant problem \cite{Arkani-Hamed2000,Neupane2011,Starkman2001,Kim2001}.

Many interesting phenomena have been found based on braneworld theories, especially the RS models and their extensions \cite{Kim2001,Neupane2011,Shiromizu2000,Davoudiasl2000,Gherghetta2000,Csaki:2004ay,Rizzo2010,
Fraser2015,Agashe2015,Dillon2016,Triyanta2014}, where the brane is assumed infinitely thin. However, the brane should have a thickness from a realistic point of view because there exists a smallest scale in a fundamental theory. When the energy corresponding to the thickness of the brane is as large as the one we are concerned about, the thickness of the brane cannot be neglected. Therefore, some thick brane models \cite{Gremm2000,DeWolfe2000,Csaki2000} have been presented, where these branes with thickness are generated by scalar fields coupled with gravity. In this thick brane scenario, the brane is explained as a domain wall where our four-dimensional universe is located. A thick brane (also called as domain wall) model is usually based on the minimal (or nonminimal) coupling between gravity and one or several bulk scalar fields \cite{Guo2015a,Bazeia2013,Afonso2006,Gremm2000,Yang2012,SouzaDutra2015,
DeWolfe2000,Bazeia2006,Liu2010c,Dzhunushaliev2010,Csaki2000,German2014,Afonso2007}. One can also use vector fields \cite{Geng2016} or pure geometry \cite{Arias2002,Barbosa-Cendejas2005,Barbosa-Cendejas2006,Liu2012b,Dzhunushaliev2010c,Zhong2016} to generate a thick brane. See Refs.~\cite{Dzhunushaliev2010,LiuZhongYang2017} for more comprehensive reviews of various thick branes.

In the thin braneworld scenario \cite{Arkani-Hamed1998,Randall1999}, matter fileds are assumed to be confined on the brane, and only gravity is free to propagate on the brane and in the bulk. However, in thick braneworld scenario, all fields (including gravity and matter fields) are in the bulk. Therefore, the zero modes of all matter fields, which stand for our four-dimensional matter fields, should be localized on the brane in order not to contradict with the low-energy experiments. Thus, localization of bulk matter fields is a significant and interesting topic in thick braneworld theories, and it is meaningful to explore how to confine matter fields on the brane in a natural way. In general, free massless scalar fields can be localized on the RS-type brane \cite{Bajc2000,Oda2001}. Free vector fields cannot be localized on the RS-type brane embedded in five-dimensional spacetime, but they can be localized on the RS brane embedded in some higher-dimensional spacetime \cite{Oda2000} or some de-sitter thick branes with finite extra dimension \cite{Liu2011} or Weyl thick branes \cite{Liu2008c}. In recent years, some new and interesting localization mechanisms have been introduced \cite{Guerrero2010,Cruz:2012kd,Chumbes2012a,Germani2012,Zhao2014,Alencar2014,Vaquera-Araujo2015,Zhao2015a} for vector fields.

Since the most elementary particles composing the matters in our four-dimensional universe are spin$-1/2$ fermions, it is very important to study localization of fermions on brane in thick braneworld theories. There are many works (see for examples Refs.~\cite{Flachi2009,Chumbes2011a,Slatyer2007a,Salvio2007,Kodama2009,Ringeval2002,
Melfo2006,Fu2011,Kehagias2001,Jardim2015,
Tahim2009,Castillo-Felisola2012a,Gogberashvili2014,Choudhury2015,Barbosa-Cendejas2015b,
Guo2015d,Cartas-Fuentevilla2016}) that study the characteristics of the fermion localization.
In the brane models where the branes are generated by one or several scalar fields, one can introduce the coupling between fermions and the background scalar fields to trap fermions on the branes.
When the background scalar field is an odd function of the extra dimension, a usual choice is the Yukawa coupling between fermions and the background scalar field  \cite{Castillo-Felisola2012a,Barbosa-Cendejas2015b,Jardim2015,Cartas-Fuentevilla2016,Flachi2009,
Chumbes2011a,Slatyer2007a,Kodama2009,Ringeval2002,Melfo2006,Fu2011,Salvio2007}.
However, when the background scalar field is an even function of the extra dimension, in order to get an effective potential function with $Z_2$ symmetry along the extra dimension, the Yukawa coupling cannot work any more.
In order to solve the problem, other coupling mechanism is needed. In the recent paper \cite{Liu2014}, the authors adopted the following derivative coupling
\begin{eqnarray}
\eta\bar{\Psi}\Gamma^M\partial_M{F(\phi)}\gamma^5\Psi,
\label{DerivativeCoupling}
\end{eqnarray}
for which the scalar field can be an odd or even function of the extra dimension. If there are even and odd background  scalars in the thick brane model, one can also consider both the above mentioned couplings at the same time \cite{Guo2015d}.

It is clear that all the previous localization mechanisms for fermions are based on the introduction of the background scalar fields. However, there are some thick brane models without scalar fields \cite{Geng2016,Arias2002,Barbosa-Cendejas2005,Barbosa-Cendejas2006,Liu2012b,Dzhunushaliev2010c,Zhong2016}. In this scenario, both mechanisms mentioned above are no longer applicable, so a new coupling mechanism should be introduced. To this end, we would like to consider the role of gravity. Gravity will always exist in various brane models since it is a geometric property of space-time. Inspired by the non-minimal coupling between gravity and other matter fields (called as geometrical coupling) introduced in localization of other matter fields \cite{Zhao2014,Alencar2014,Jardim2015,Alencar2016}, we naturally think of non-minimal coupling between gravity and fermions. One of the most intuitive thoughts is to consider that the scalar curvature $R$ couples to fermion fields, namely the Yukawa-like geometrical coupling $\eta R \bar{\Psi}\Psi$ (the similar couplings have been considered for vector fields \cite{Zhao2014,Alencar2014,Alencar2016} and Elko spinor fields \cite{Jardim2015} \footnote{Elko spinor fields are eigenspinors of the charge conjugation operator, and were introduced by Ahluwalia-Khalilova and Grumiller in 2005 \cite{{Ahluwalia-Khalilova2005}}. They are spin 1/2 fermionic quantum fields with mass dimension one in four dimensions.} ). However, in order to get an effective potential with $Z_2$ symmetry along the extra dimension, and because the scalar curvature is an even function of the extra dimension, we adopt the derivative geometrical coupling between gravity and fermions, like the one given in \eqref{DerivativeCoupling}, rather than the Yukawa-like coupling. This is inspired by the paper \cite{Liu2014}.
In this paper, we will focus on the fermion localization and mass spectrum by considering the coupling  $\eta\bar{\Psi}\Gamma^M\partial_M{F(R)} \gamma^5\Psi$. The new fermion localization mechanism depends on the brane structure and the form of $F(R)$.  We will consider some specific forms of $F(R)$ for three brane models to localize fermions on the branes. It will be shown that three kinds of well-known effective potentials (volcano-like, P$\ddot{\text{o}}$schl-Teller(PT)-like and harmonic-like potentials) for the left- and right-chiral fermion KK modes can be obtained.

The paper is organized as follows.
In Sec. \ref{Section2}, we mainly introduce a new localization mechanism with a derivative geometrical coupling and derive the effective Dirac action of the fermion KK modes including the four-dimensional massless and massive Dirac fermions. We also obtain the effective potentials and the orthonormality conditions for the fermion KK modes.
In Sec. \ref{Section3}, we study three kinds of thick brane models as examples.
The first model is about a sine-Gordon brane generated by a single scalar field. We investigate three different forms of $F(R)$ corresponding to three typical effective potentials and obtain the corresponding mass spectra for this model.
In the second model, we also study a single-scalar-field-generated thick brane and calculate the mass spectrum of a fermion with $F(R)=R$.
In particular, in the last model, we consider the case of a pure geometric braneworld.
Finally, a brief discussion and conclusion is presented in Sec. \ref{Section4}.

\section{Localization mechanism with a derivative geometrical coupling} \label{Section2}

In order to localize spin-$\frac{1}{2}$ fermions on a brane with codimension one, one usually needs to introduce some localization mechanisms.
In this section we introduce a new coupling between a spin-$\frac{1}{2}$ fermion $\Psi$ and the scalar curvature $R$ of the five-dimensional spacetime $R$: $\eta\bar{\Psi}\Gamma^M\partial_M{F(R)}\gamma^5\Psi$, where $F(R)$ is a function of $R$. The corresponding Dirac action for a massless spin-$\frac{1}{2}$ fermion coupled to the scalar curvature $R$ is
\begin{eqnarray}
      S_{1/2}=\int d^5 x \sqrt{-g}\left[\bar \Psi \Gamma^M (\partial _M + \omega _M)\Psi
              +\eta \bar \Psi \Gamma^M \partial_M F(R)\gamma ^5 \Psi \right] ,
              \label{the five-dimensional action}
\end{eqnarray}
where $\eta$ is a coupling constant and the spin connection $\omega_M$ is given by
\begin{eqnarray}
\omega_{M}
 = \frac{1}{4}\omega^{\bar{M}\bar{N}}_{M}
         \Gamma_{\bar{M}}\Gamma_{\bar{N}}
\label{covariant derivative}
\end{eqnarray}
with
\begin{eqnarray}
\omega^{\bar{M}\bar{N}}_{M}
 &=&\frac{1}{2}E^{N\bar{M}}(\partial_{M}E^{\bar{N}}_{N}
         -\partial_{N}E^{\bar{N}}_{M})  \nonumber\\
 &-& \frac{1}{2}E^{N\bar{N}}(\partial_{M}E^{\bar{M}}_{N}
         -\partial_{N}E^{\bar{M}}_{M})\nonumber\\
 &-& \frac{1}{2}E^{P\bar{M}}E^{Q\bar{N}} E^{\bar{R}}_{M}
   (\partial_{P}E_{Q\bar{R}}-\partial_{Q}E_{P\bar{R}})\,.
   \label{SpinConnection}
\end{eqnarray}
In five-dimensional spacetime, the anticommutation relation of the gamma matrices is $\{\Gamma^{M},\Gamma^{N}\}=2g^{MN}$, where $\Gamma^{M}=E^{M}_{\bar{M}}\Gamma^{\bar{M}}$ with $E^{M}_{\bar{M}}$ and $\Gamma^{\bar{M}}$ the vielbein and gamma matrix of the flat five-dimensional spacetime, respectively.
The capital Latin letters $M,N...$ denote general five-dimensional coordinate indices; the indices $\bar{M},\bar{N}...$ denote the local Lorentz frame indices.

The five-dimensional line element of a Minkowski brane takes the form
\begin{equation}
  ds^{2}_{5}=e^{2A(y)}\eta_{\mu\nu}dx^{\mu}dx^{\nu}+dy^{2} .
  \label{Metric}
\end{equation}
Here $e^{2A(y)}$ is the warp factor, $x^\mu$ are coordinates for the usual four dimensions, $y$ denotes the extra-dimensional coordinate, and $\eta_{\mu\nu}$ is the induced metric on the brane.
Through the conformal coordinate transformation
\begin{equation}
dz=e^{-A(y)}dy, \label{coordinateTransformation}
\end{equation}
the line element (\ref{Metric}) can be rewritten as
\begin{equation}
ds^{2}_{5}=e^{2A(z)}(\eta_{\mu\nu}dx^{\mu}dx^{\nu}+dz^{2}).
\label{conformalFlatMetric}
\end{equation}
The components of the spin connection are given by  $\omega_{\mu}=\frac{1}{2}({\partial_z}A(z))\gamma_{\mu}\gamma_{5}$
and $\omega_5=0$ for the conformally flat metric (\ref{conformalFlatMetric}). Then the equation of motion of the five-dimensional Dirac fermion can be derived as
\begin{eqnarray}
\Big[\gamma^{\mu}\partial_{\mu}
     +\gamma^5 \big(\partial_z+2{\partial_z}A(z)\big)
     +\eta \partial_z F(R) \Big]\Psi=0.\label{DiracEq1}
\end{eqnarray}
Now we make the chiral decomposition for $\Psi(x,z)$
\begin{eqnarray}
 \Psi(x,z) =
      \sum_{n}\psi_{Ln}(x)f_{Ln}(z) e^{-2A(z)}
       + \sum_{n}\psi_{Rn}(x)f_{Rn}(z) e^{-2A(z)},
\label{the general chiral decomposition}
\end{eqnarray}
where $\psi_{Ln}$ and $\psi_{Rn}$ are the left- and right-chiral components of the four-dimensional effective Dirac field $\psi$, respectively, with $\psi_{Ln}(x)=-\gamma^{5}\psi_{Ln}(x)$ and $\psi_{Rn}(x)=\gamma^{5}\psi_{Rn}(x)$, and they satisfy the four-dimensional massive Dirac equations $\gamma^{\mu}\partial_{\mu}\psi_{Ln}(x)=\mu_{n}\psi_{Rn}(x)$ and
$\gamma^{\mu}\partial_{\mu}\psi_{Rn}(x)=\mu_{n}\psi_{Ln}(x)$.
By substituting the decomposition (\ref{the general chiral decomposition}) into Eq. (\ref{DiracEq1}), we can get the equations of motion for the left- and right-chiral fermion KK modes $f_{Ln}(z)$ and $f_{Rn}(z)$:
\begin{eqnarray}
&&[\partial_z-\eta\partial_z F]f_{Ln}(z)=+\mu_n f_{Rn}(z), \label{spinorCoupledEqs1}\\
&&[\partial_z+\eta\partial_z F]f_{Rn}(z)=-\mu_n f_{Ln}(z). \label{spinorCoupledEqs2}
\end{eqnarray}
The above two equations can also be rewritten as
\begin{subequations}\label{SuperSymmetricForm}
\begin{eqnarray}
&&U^{\dagger}U f_{Ln}(z)=\mu_n^2 f_{Ln}(z), \\
&&U U^{\dagger}f_{Rn}(z)=\mu_n^2 f_{Rn}(z),
\end{eqnarray}
\end{subequations}
where $U\equiv\partial_z-\eta\partial_z F$. These equations can be written as the Schr\"{o}dinger-like equations
\begin{subequations}\label{Scheq}
\begin{eqnarray}
 [-\partial_{z}^{2}+V_{L}(z)]f_{L}(z) &=& \mu_{n}^{2}f_{L}(z) \,,
    \\ \label{ScheqLeft}
 [-\partial_{z}^{2}+V_{R}(z)]f_{R}(z) &=& \mu_{n}^{2}f_{R}(z) \,.
       \label{ScheqRight}
\end{eqnarray}
\end{subequations}
The effective potentials $V_{L,R}(z)$ of the fermion KK modes $f_{L,R}(z)$ read
\begin{eqnarray}
V_{L,R}(z)=\big(\eta\partial_z F\big)^2
        \pm \partial_{z}\big(\eta\partial_z F\big).  \label{VzLR}
\end{eqnarray}
In order to reduce the five-dimensional Dirac action to the effective Dirac action of the four-dimensional massless and massive Dirac fermions:
\begin{eqnarray}
      S_{1/2}=\sum_n\int d^4 x \left[\bar \psi_n \gamma^{\mu}\partial_{\mu}\psi_n-{\mu_n}\bar\psi_n\psi_n  \right] ,
              \label{the five-dimensional action}
\end{eqnarray}
the KK modes $f_{L,R}(z)$  should satisfy the following orthonormality conditions
\begin{eqnarray}
   \int_{-\infty}^{+\infty} f_{Lm}f_{Ln}dz &=& \int_{-\infty}^{+\infty} f_{Rm}f_{Rn}dz=\delta_{mn}, \nonumber \\
   \int_{-\infty}^{+\infty} f_{Lm}f_{Rn}dz &=& 0.
  \label{orthonormality conditions}
\end{eqnarray}
From Eqs. (\ref{spinorCoupledEqs1}) and (\ref{spinorCoupledEqs2}), the left- and right-chiral fermion zero modes are
\begin{eqnarray}
 f_{L0,R0}(z) \propto
  \exp\bigg(\pm
   \int_{0}^{z} d\bar{z}
        \eta \partial_{\bar{z}} F
         \bigg)
  = \exp\big(\pm\eta  F \big).     \label{fL0CaseI}
\end{eqnarray}
So the normalization condition is
\begin{eqnarray}
 \int_{-\infty}^\infty\exp\big(\pm2\eta  F \big) dz < \infty.
       \label{fL0CaseI}
\end{eqnarray}
It can be seen that if the conformal coordinate $z$ is infinite, at most one of the left- and right-chiral zero modes can be localized on the brane.

\section{Localization of fermions on branes}\label{Section3}
Next, we mainly investigate the localization problem of a spin-$1/2$ fermion on a thick brane with the new coupling $\eta\bar{\Psi}\Gamma^M\partial_M{F(R)}\gamma^5\Psi$ proposed in the previous section. We take three brane scenarios as examples. In these examples, we consider two kinds of single-scalar-generated thick branes with different scalar potentials and a pure geometric thick brane. In the first brane model, we construct three kinds of $F(R)$ that generate three well-known effective potentials. In the second model, we mainly study the case of $F(R)=R$. For the last example, we briefly study the case of a pure geometric thick brane.

\subsection{Brane model I}

We first consider a sine-Gordon brane generated by a single scalar field in a five-dimensional spacetime. The corresponding action for this system reads as
   \begin{eqnarray}
      S = \int d^4x dy \sqrt{-g}\left[ \frac{1}{4} R -
           \frac{1}{2}(\partial\phi)^2 - V(\phi) \right],
   \end{eqnarray}
where $R$ is the five-dimentional scalar curvature.
The metric ansatz is also (\ref{Metric}).
The field equations are derived as follows:
   \begin{eqnarray}
      \phi^{\prime\prime}+4A^\prime \phi^\prime &=&
       \frac{\partial V(\phi)}{\partial\phi}, \nonumber\\
       A^{\prime\prime} &=& -\frac{2}{3} \phi^{\prime 2}, \nonumber\\
           A^{\prime 2} &=& -\frac{1}{3}V(\phi)+\frac{1}{6}\phi^{\prime 2}\label{fieldequation},
   \end{eqnarray}
where the prime denotes the derivation with respect to $y$, besides, $\phi$ and $A$ only depend on $y$.

A brane solution was obtained in Ref.~\cite{Gremm2000} for the sine-Gordon potential:
  \begin{eqnarray}
     V(\phi) &=& \frac{3 bc^2}{8}
     \left[ (1-4b)-(1+4b)\cos(\sqrt{\frac{8}{3b}}\phi)\right], \\
     \label{v1}
     A(y) &=& -b \ln\left(2 \cosh(c y) \right), \label{warpfactor}\\
     \phi(y) &=& \sqrt{6b}\, {\rm arctan}
	 \left( \tanh( \frac{c y}{2})\right)\label{phi1},
  \end{eqnarray}
where both the parameters $b$ and $c$ are positive constants. In this paper, we consider the case of $b=1$, for which the conformal transformation (\ref{coordinateTransformation}) is given by  $z = \int e^{-A(y)} dy =\frac{2 \sinh (c y)}{c} $, and the warp factor and the scalar curvature in the conformal coordinate read $A(z) = -\ln( \sqrt{4 + c^2 z^2} )$ and $R=4c^2\left(\frac{28}{4+c^2 z^2}-5\right)$, respectively.

Now we can apply the new localization mechanism to localize fermions on the brane by some suitable choices of the function $F(R)$.
For different forms of $F(R)$, we will get different effective potentials for the left- and ritht-chiral fermion KK modes.

\subsubsection{$F(R)=\ln ^q\left(\frac{20 c^2+R}{52 c^2-3R}\right)$}

First, we choose $F(R)=\ln ^q\left(\frac{20 c^2+R}{52 c^2-3R}\right)$ with $q$ a positive integer. The corresponding effetive potentials are
  \begin{eqnarray}
    V_{L,R}(z)&=&\frac{2 q \eta~ c^2 \ln^{(q-2)}\left(\frac{1}{1+c^2z^2}\right)}{\left(1+c^2 z^2\right)^2}\nonumber\\
    && \times\left[\pm\Big(2 (q-1)c^2 z^2-\left(c^2z^2-1\right) \ln({1+c^2z^2})\Big)+2 q \eta ~c^2 z^2  \ln^q(\frac{1}{1+c^2 z^2})\right].\label{Vtype1}
  \end{eqnarray}
The values of $V_L(z)$ and $V_R(z)$ at $z=0$ and $z\rightarrow \pm \infty $ read
\begin{eqnarray}
    V_{L,R}(0)&=&\left\{
                \begin{aligned}
                 \mp2c^2\eta,~ q=1 \\
                  0, ~q\geq2\\
                \end{aligned}
                \right. \nonumber \\
    V_{L,R}(\pm \infty)&=&0.
\end{eqnarray}

\begin{figure}[htb]
\begin{center}
\subfigure[$c=1, q=1, \eta=2$]{
\includegraphics[width=5cm,height=3cm]{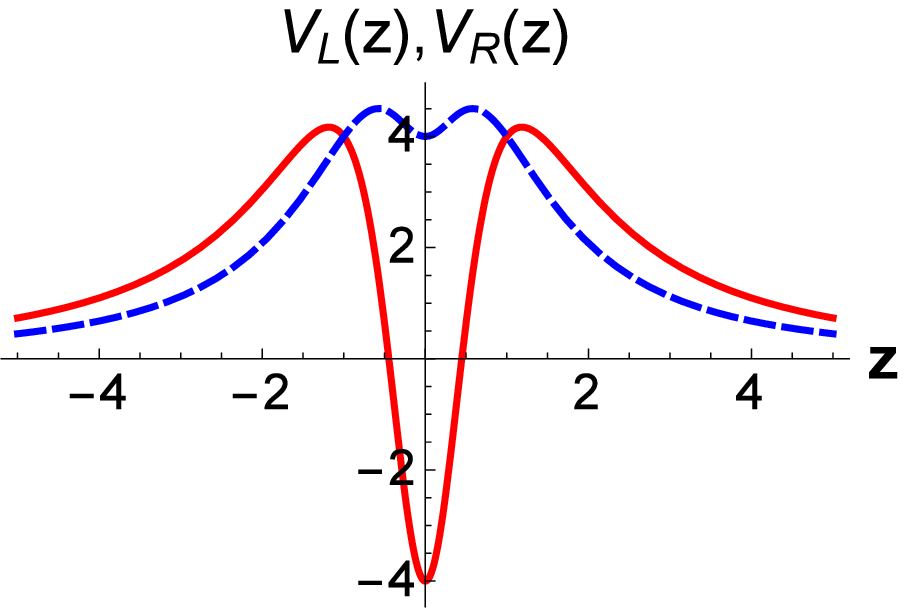}}
\subfigure[$c=1, q=2, \eta=1$]{
\includegraphics[width=5cm,height=3cm]{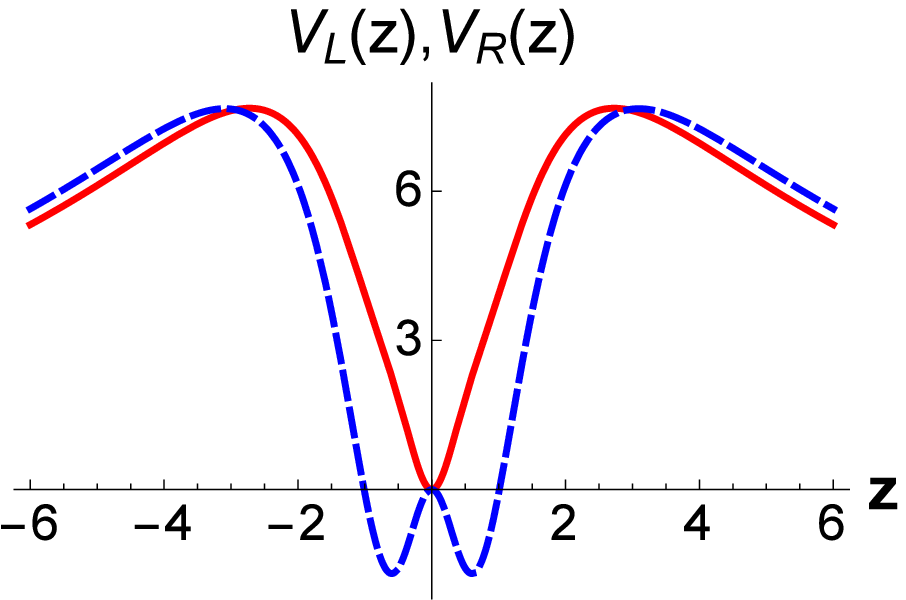}}
\subfigure[$c=1, q=3, \eta=0.2$]{
\includegraphics[width=5cm,height=3cm]{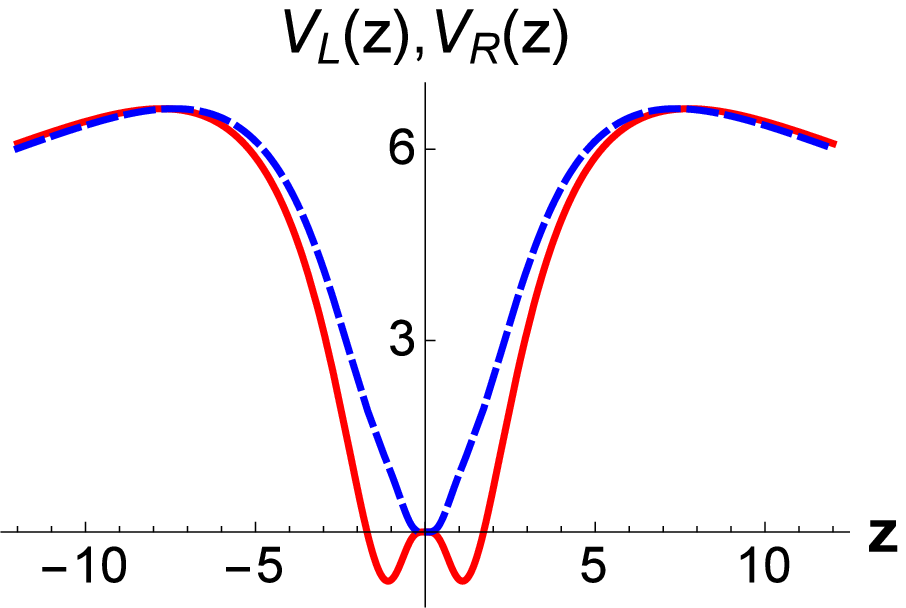}}
\end{center}
\caption{
Plots of the effective potentials $V_{L}(z)$ (red curves) and $V_{R}(z)$ (blue dashed curves) in (\ref{Vtype1}) with the coupling $F(R)=\ln ^q\left(\frac{20 c^2+R}{52 c^2-3R}\right)$ for $c=1$ and different values of $q$ and $\eta$.
}
\label{VLVR1-1}
\end{figure}

\begin{figure}[htb]
\begin{center}
\subfigure[$c=1, q=1$]{
\includegraphics[width=5cm,height=3cm]{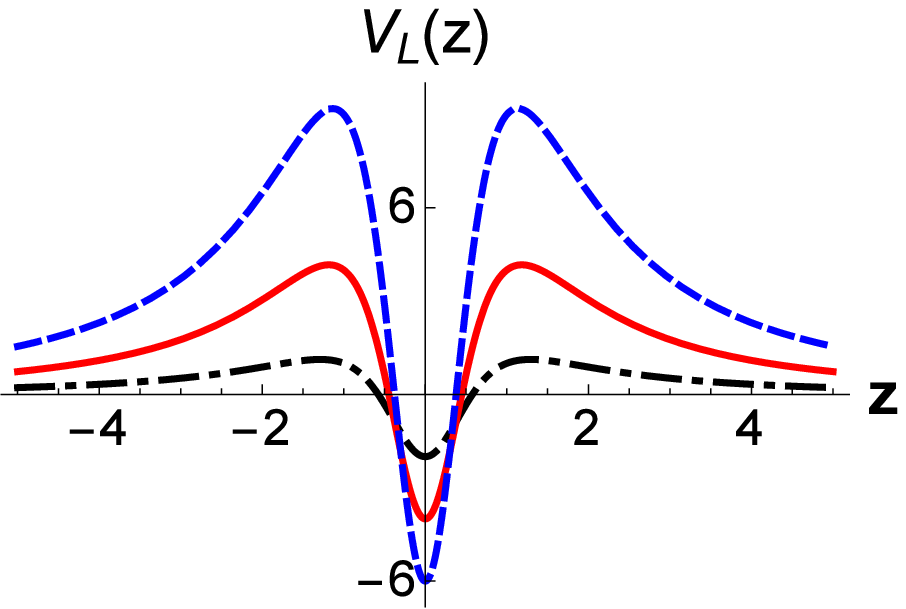}}
\subfigure[$c=1, q=1$]{
\includegraphics[width=5cm,height=3cm]{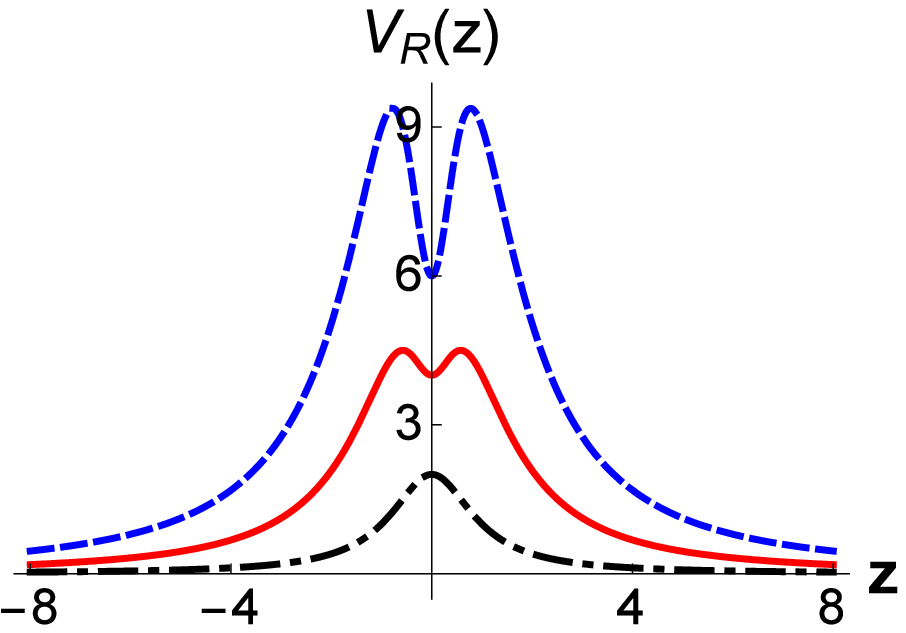}}\\
\subfigure[$c=1, q=2$]{
\includegraphics[width=5cm,height=3cm]{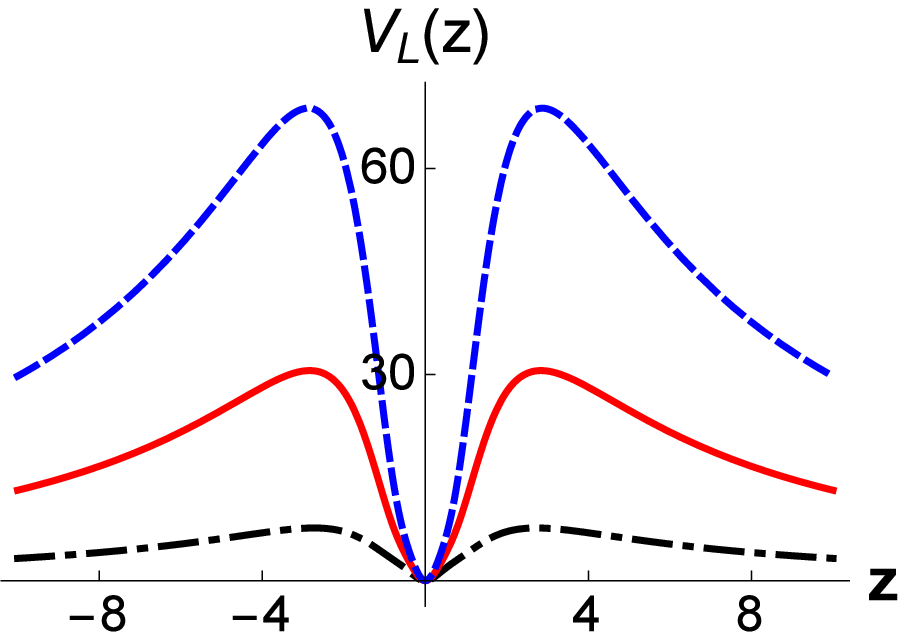}}
\subfigure[$c=1, q=2$]{
\includegraphics[width=5cm,height=3cm]{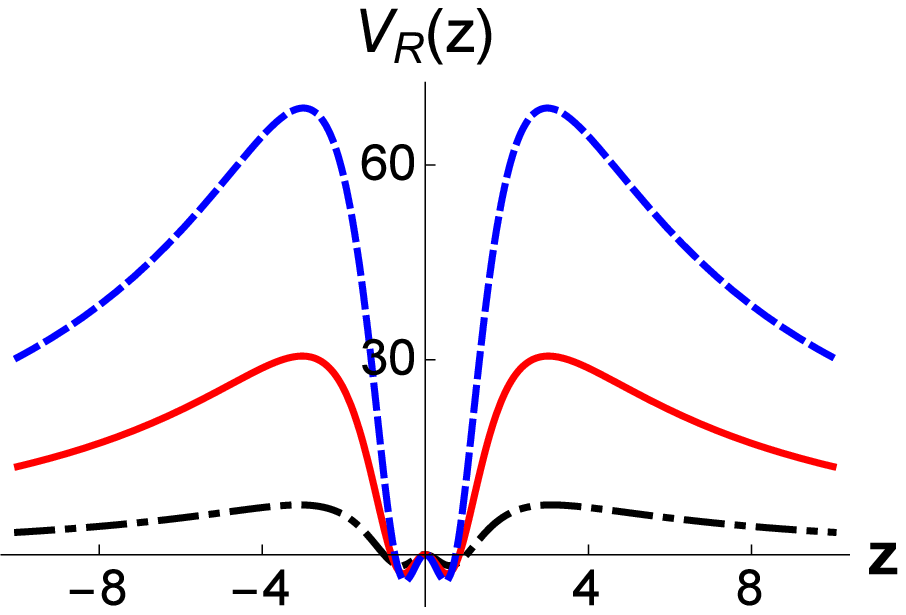}}
\end{center}
\caption{
 Plots of the effective potentials in (\ref{Vtype1}) with the coupling $F(R)=\ln ^q\left(\frac{20 c^2+R}{52 c^2-3R}\right)$  for $c=1$, $q = 1,~2$ and different values of $\eta$. The
parameter $\eta$ is  set to $\eta =1 $ (black dot dashed line), 2 (red line), and 3 (blue dashed line).
}
\label{VLVR1-2}
\end{figure}

When the coupling constant $\eta$ converts to $-\eta$, as a result, the left-chiral effective potential interchanges with the right-chiral one. Hence, we just consider the case of positive coupling constant $\eta$. When $z$ approximates to infinity, both the values of $V_L(z)$ and $V_R(z)$ vanish.  For the case of $q=1$, the values of $V_L(z)$ and $V_R(z)$ at $z=0$ are $-2c^2\eta$ and $2c^2\eta$, respectively. For the case of $q\ge 2$, both $V_L(0)$ and $V_R(0)$ are zero. Plots of the different effective potentials are shown in Fig.~\ref{VLVR1-1}, which shows that the effective potentials have the shape of the well-known volcano-like one. For odd (even) $q$, the left-chiral (right-chiral)  effective potential has a potential well, which indicates that the left-chiral fermion zero mode could be localized on the brane. For the case of $q\ge 2$, there is a double well for the left- or right-chiral effective potential. On the other hand,
for the positive $\eta$ and fixed $q$ and  $c$, the depth of the potential wells increase with $\eta$, which can be seen in Fig.~\ref{VLVR1-2}.

The solutions of the fermion zero modes are given by
\begin{eqnarray}
    f_{L_0,R_0}\propto \exp \left[\pm \eta ~ \text{ln}^q\left(\frac{1}{1+c^2z^2}\right)\right].
\end{eqnarray}
In order to localize the zero modes on the brane, they should satisfy the normalization condition
\begin{eqnarray}
 \int_{-\infty}^\infty |f_{L_0,R_0}|^2 dz < \infty,
       \label{fL0CaseI}
\end{eqnarray}
where the coupling parameter $\eta$ is positive.
When $q = 1$, the integrand function becomes $(1+c^2z^2)^{\mp2\eta}$, so only if $\eta > 1/4$, the left-chiral zero mode can be localized on the brane. It can be seen from the above formula (\ref{fL0CaseI}) that for any positive odd (even) $q$ with $q\geq2$ only the left-chiral (right-chiral) zero mode can be localized on the brane with $\eta >0$, because the integrand function
\begin{eqnarray}
|f_{L_0,R_0}|^2
  &=& \Big(\frac{1}{1+c^2z^2}\Big)^{\pm2\eta \left(\ln \frac{1}{1+c^2z^2}\right)^{q-1}} \nonumber \\
  &\rightarrow & \Big(\frac{1}{cz}\Big)^{\pm4\eta \left(\ln \frac{1}{c^2z^2}\right)^{q-1}}\nonumber \\
  &\rightarrow & \Big(\frac{1}{cz}\Big)^{\pm4\eta(-\infty)^{q-1}}
\end{eqnarray}
when $z\rightarrow \infty$.

\subsubsection{$F(R)=\ln ^q\left(\text{sech}(\frac{2 \sqrt{8 c^2-R}}{ \sqrt{20 c^2+R}} )\right)$}

\begin{figure}[htb]
\begin{center}
\subfigure[$c=1, q=1, \eta=0.7$]{
\includegraphics[width=5cm,height=3cm]{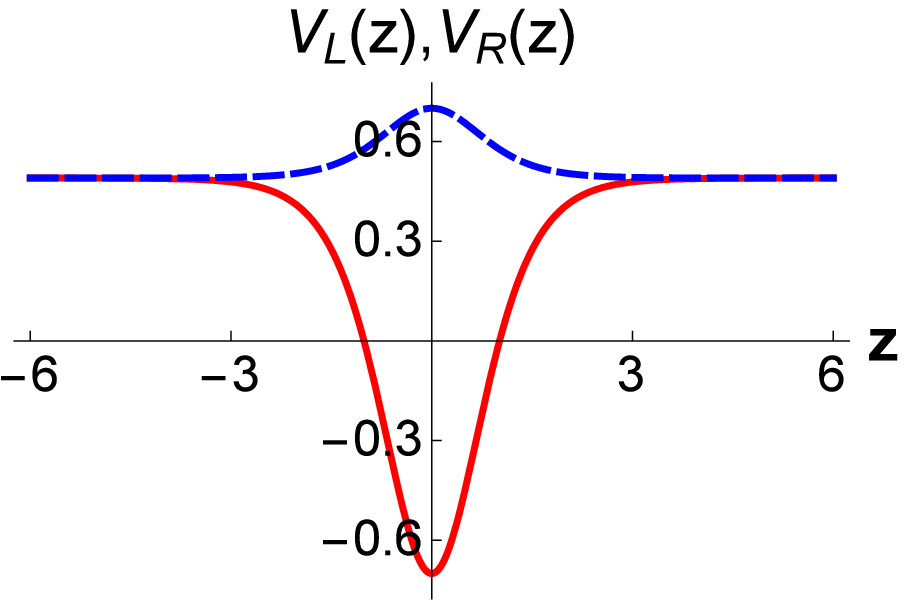}}
\subfigure[$c=1, q=1, \eta=2.5$]{
\includegraphics[width=5cm,height=3cm]{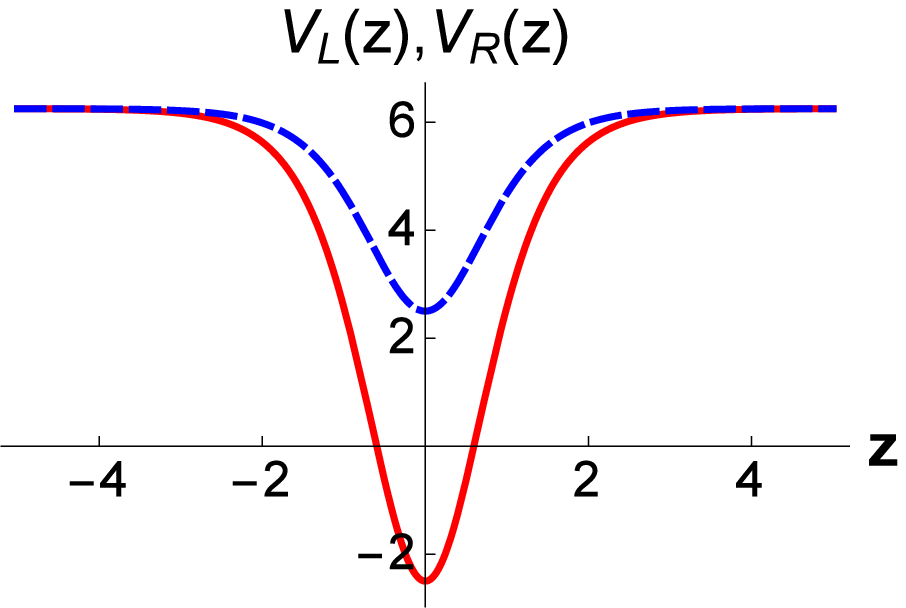}}\\
\subfigure[$c=1, q=2, \eta=0.3$]{
\includegraphics[width=5cm,height=3cm]{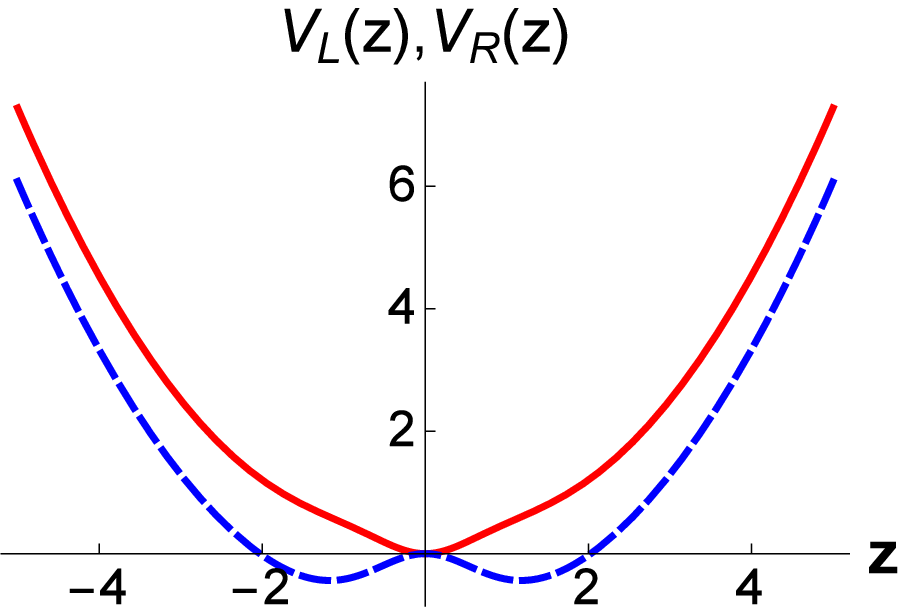}}
\subfigure[$c=1, q=3, \eta=0.05$]{
\includegraphics[width=5cm,height=3cm]{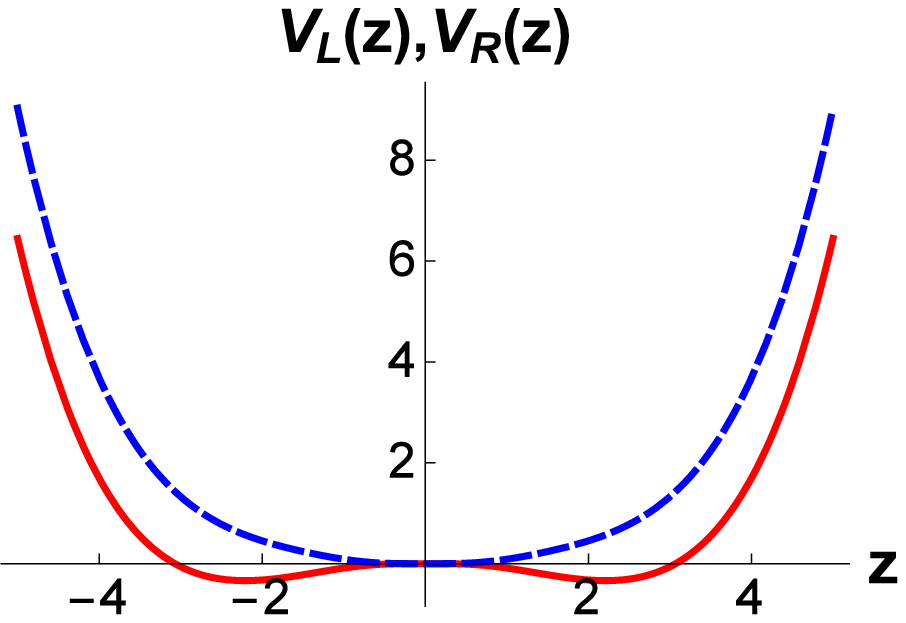}}\\
\end{center}
\caption{
Plots of the effective potentials $V_{L}(z)$ (red curves) and $V_{R}(z)$ (blue dashed curves) in (\ref{Vtype2}) with the coupling $F(R)=\ln ^q\left(\text{sech}(\frac{2 \sqrt{8 c^2-R}}{ \sqrt{20 c^2+R}} )\right)$ for $c=1$ and different values of $q$ and $\eta$.
}
\label{VLVR2-1}
\end{figure}

Next we consider
$F(R)=\ln ^q\left(\text{sech}(\frac{2 \sqrt{8 c^2-R}}{ \sqrt{20 c^2+R}} )\right)$, for which the effective potentials read
  \begin{eqnarray}
    V_{L,R}(z)&=&  q\eta c^2\ln ^{q-2}(\text{sech}(cz)) \nonumber\\
    && \times\Big\{ \Big[\pm(q-1)+ q\eta\ln ^q(\text{sech}(cz))\Big]\tanh ^2(cz)
    \mp\text{sech}^2(cz) \ln \left(\text{sech}(cz)\right)\Big\}.\label{Vtype2}
  \end{eqnarray}
We can also get
\begin{eqnarray}
    V_{L,R}(0)&=&\left\{
                \begin{array}{cc}
                 \mp c^2\eta,&~ q=1 \\
                  0, &~q\geq2\\
                \end{array}
                \right.  \\
    V_{L,R}(\pm \infty)&=&\left\{
                \begin{array}{cc}
                 c^2\eta^2,&~ q=1 \\
                  +\infty. &~q\geq2\\
                \end{array}
                \right.
\end{eqnarray}
The left- and right-chiral effective potentials for different $q$ and $\eta$ are plotted in  Fig.~\ref{VLVR2-1}.

\begin{figure}[htb]
\begin{center}
\subfigure[$c=1, q=1, \eta=5$]{
\includegraphics[width=5cm,height=3cm]{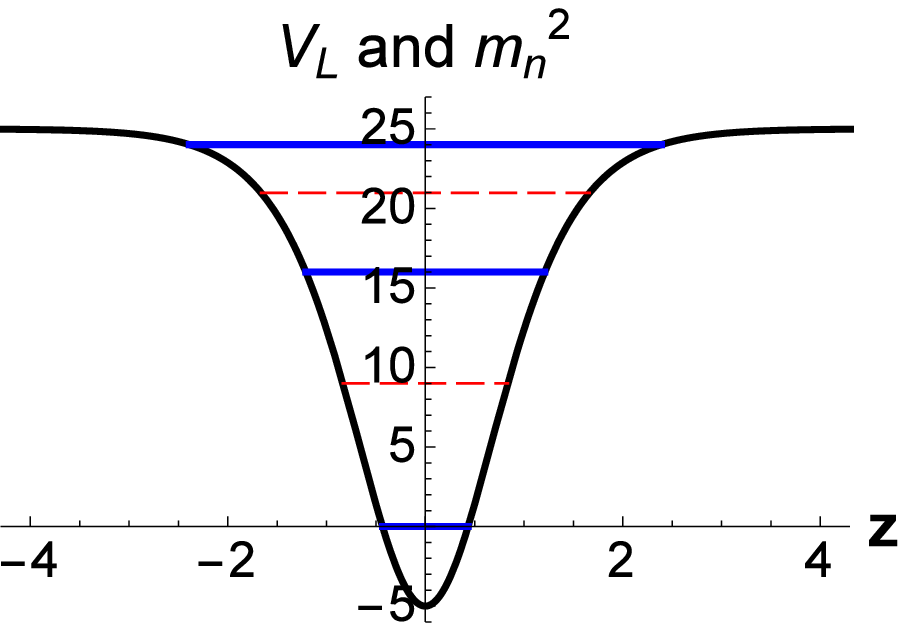}}
\subfigure[$c=1, q=1, \eta=5$]{
\includegraphics[width=5cm,height=3cm]{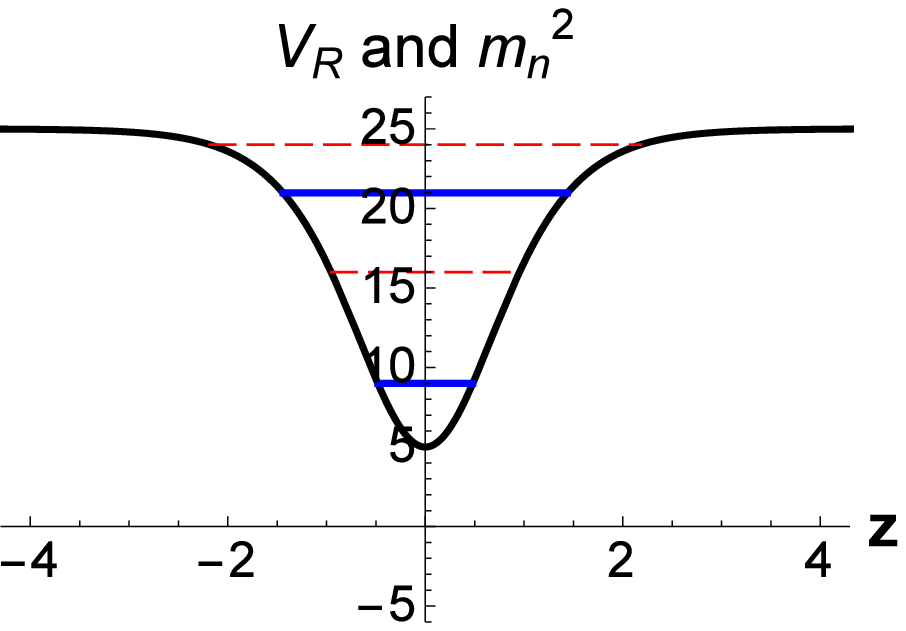}}\\
\subfigure[$c=1, q=1, \eta=10$]{
\includegraphics[width=5cm,height=3cm]{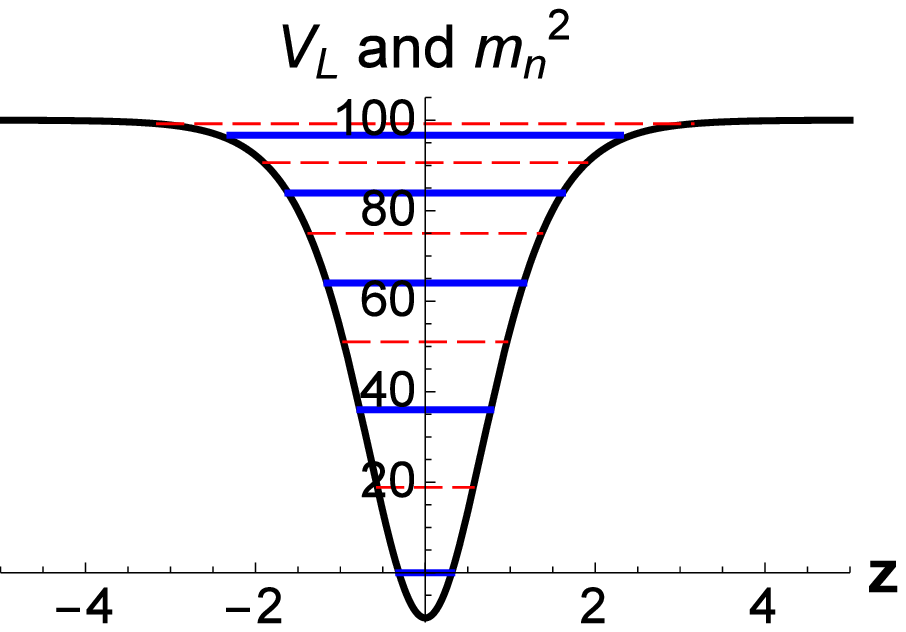}}
\subfigure[$c=1, q=1, \eta=10$]{
\includegraphics[width=5cm,height=3cm]{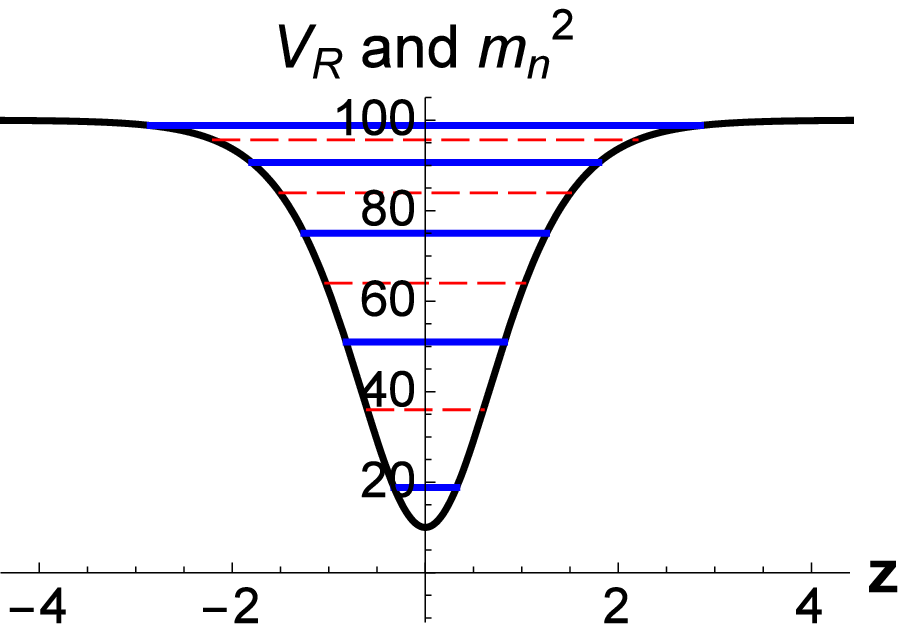}}\\
\end{center}
\caption{
The left- and right-chiral effective potentials $V_{L,R}(z)$ and the mass spectra $m_n^2$ in (\ref{Vtype2}) with even parity (blue curves) and odd  parity (red dashed curves) with the coupling $F(R)=\ln ^q\left(\text{sech}(\frac{2 \sqrt{8 c^2-R}}{ \sqrt{20 c^2+R}} )\right)$ for $c=1$ and $q=1$. The parameter $\eta$
is set to $\eta=5$ and $\eta=10$, respectively.
}
\label{PTmass}
\end{figure}

Also, we only consider the case of positive coupling parameter $\eta$.
It can be seen that in the case of $q=1$, we have $V_L(0)<0$ and $V_R(0)>0$. Since $V_{L}(0)<0$ and the value of $V_{L}$ is a positive constant at infinity (this kind of potential is called P$\ddot{\text{o}}$schl-Teller(PT)-like), there exists a mass gap in the mass spectrum. The left-chiral fermion has discrete mass spectrum with $m_n^2< c^2\eta^2$, and continuous mass spectrum with $m_n^2> c^2\eta^2$. However, for the right-chiral fermion, there may be a mass gap when $\eta>1$. Furthermore, the right-chiral fermion cannot be localized on the brane when $0<\eta \leq 1$,
and at the same time there is only one bound state with zero mass for the corresponding left-chiral fermion.
We can obtain the mass spectrum with a numerical method. For example, the mass spectra $m^2_{L_n}$ of the left-chiral fermions ($q=1,c=1$) are
 \begin{eqnarray}
 m_{L_n}^2 &=& (0,2.96)\cup(4,\infty),
     ~\text{for}~\eta=2,\\
     m_{L_n}^2 &=& (0,4.93,7.95)\cup(9,\infty),
     ~\text{for}~\eta=3,\\
     m_{L_n}^2 &=& (0,9.00,16.00,20.98,24.01)\cup(25,\infty),
     ~\text{for}~\eta=5,\\
     m_{L_n}^2 &=& (0,18.84,36.00,50.98,64.00,75.00,83.91,90.63,96.65,99.20)\nonumber\\
     &&\cup(100,\infty),
     ~\text{for}~\eta=10.
  \end{eqnarray}
Here, we plot the effective potentials and the mass spectra of the fermion KK modes in the case of $q=1$ and $c=1$ in Fig. \ref{PTmass}. This also validates the view we mentioned earlier that the ground state of the left-chiral fermion is a massless zero mode, while it is a massive KK mode for the right-chiral fermion. We obtain a massless left-chiral fermion and some massive left- and right-chiral KK mode fermions with the same mass. It can also be shown that the mass of the first massive bound state and the number of the bound states increase as $\eta$ increases, which is because that $V_{L}(\pm \infty)=c^2 \eta^2$ and the potential depth of $V_{L}(z)$ increases with $\eta$. Besides, the discrete mass spectrum becomes more density with the increase of the mass.
Moreover, we can get the wave functions of the KK bound states with odd parity and even parity:
\begin{eqnarray}
                f_{L_n}(z)&=&\left\{
                \begin{array}{ll}
                {C_1}\Big[P_{\eta }^{\zeta}\big(\tanh (z)\big)-\frac{P_{\eta +1}^{\zeta}(0)}{Q_{\eta +1}^{\zeta}(0)} Q_{\eta }^{\zeta}\big(\tanh (z)\big) \Big], &~~~~~~\text{even parity}\\
                 {C_2}\Big[P_{\eta }^{\zeta}\big(\tanh (z)\big)-\frac{P_{\eta }^{\zeta}\left(0\right)}{Q_{\eta }^{\zeta}\left(0\right)} Q_{\eta }^{\zeta}\big(\tanh (z)\big)\Big],&~~~~~~\text{odd ~parity}  \\
                \end{array}
                \right.  \\
                f_{R_n}(z)&=&\left\{
                \begin{array}{ll}
                {C_3}\Big[ P_{\eta -1}^{\zeta}\big(\tanh (z)\big)-\frac{P_{\eta }^{\zeta}(0)}{Q_{\eta }^{\zeta}(0)} Q_{\eta -1}^{\zeta}\big(\tanh (z)\big)\Big],&~\text{even parity}  \\
                {C_4}\Big[ P_{\eta -1}^{\zeta}\big(\tanh (z)\big)-\frac{P_{\eta -1}^{\zeta}(0)}{Q_{\eta -1}^{\zeta}(0)}Q_{\eta -1}^{\zeta}\big(\tanh (z)\big)\Big], &~\text{odd ~parity}\\
                \end{array}
                \right.
\end{eqnarray}
where, $\zeta\equiv {\sqrt{\eta ^2-m_n^2}}$,  $P_{\eta}^{\zeta}(w)$ and $Q_{\eta }^{\zeta}(w)$ represent the associated Legendre functions of the first and second kinds, respectively, and $C_{1,2,3,4}$ are normalization constants.

\begin{figure}[htb]
\begin{center}
\subfigure[$m^2_{L_n}$]{
\includegraphics[width=6cm,height=3.6cm]{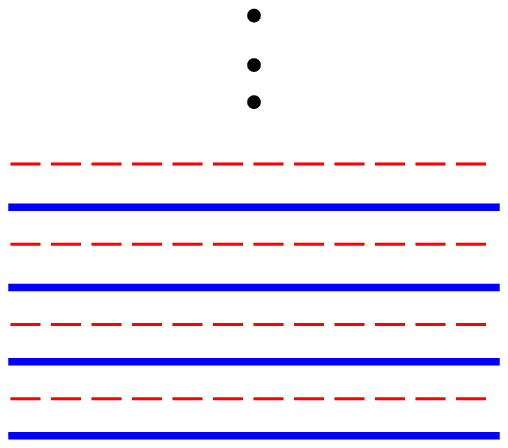}}
\subfigure[$m^2_{R_n}$]{
\includegraphics[width=6cm,height=3.6cm]{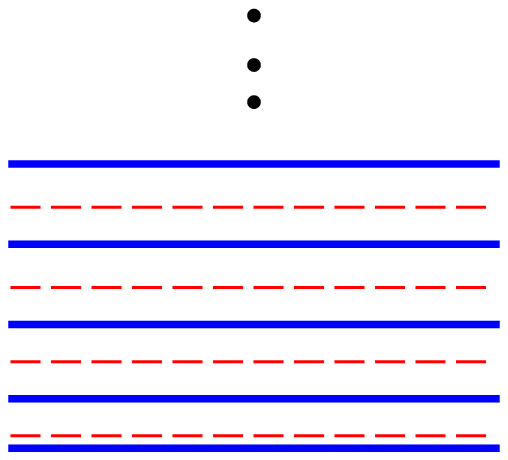}}
\end{center}
\caption{
The mass spectra $m^2_{L_n,R_n}$ of the left- and right-chiral fermions with even parity (blue curves) and odd  parity (red dashed curves) in (\ref{Vtype2}) with the coupling $F(R)=\ln ^q\left(\text{sech}(\frac{2 \sqrt{8 c^2-R}}{ \sqrt{20 c^2+R}} )\right)$ for $c=1, q=2$ and $\eta=1$.
}
\label{mass2}
\end{figure}

For $q\geq2$, there are discrete mass spectra for both the left- and right-chiral fermions. When $q$ is odd, the left-chiral effective potential $V_L$ has a double well with negative values around the brane, which could cause the localization of the zero mode, while the right-chiral effective potential $V_R$ is an infinite potential well without negative values and so the right-chiral zero mode cannot be localized. For even $q$, the shapes of $V_L$ and $V_R$ are opposite. We plot the mass spectra $m^2_{L_n,R_n}$ of the lower bound KK modes  of the left- and right-chiral fermions in the case of $q=2$ in Fig. \ref{mass2} and the mass spectra are listed as follows:
\begin{eqnarray}
 m_{L_n}^2 &=& (~~~1,4,7,10,13,16.5, 19.5,23,\cdots),
     ~\text{for}~\eta=1,~q=2,\\
     m_{R_n}^2 &=& (0,1,4,7,10,13,16.5, 19.5,23,\cdots),
     ~\text{for}~\eta=1,~q=2.
\end{eqnarray}
It is shown that only the right-chiral fermion zero mode can be localized on the brane, and the left- and right-chiral fermions have the same KK bound states with opposite parity.

\begin{figure}[htb]
\begin{center}
\subfigure[$c=1, q=1$]{
\includegraphics[width=5cm,height=3cm]{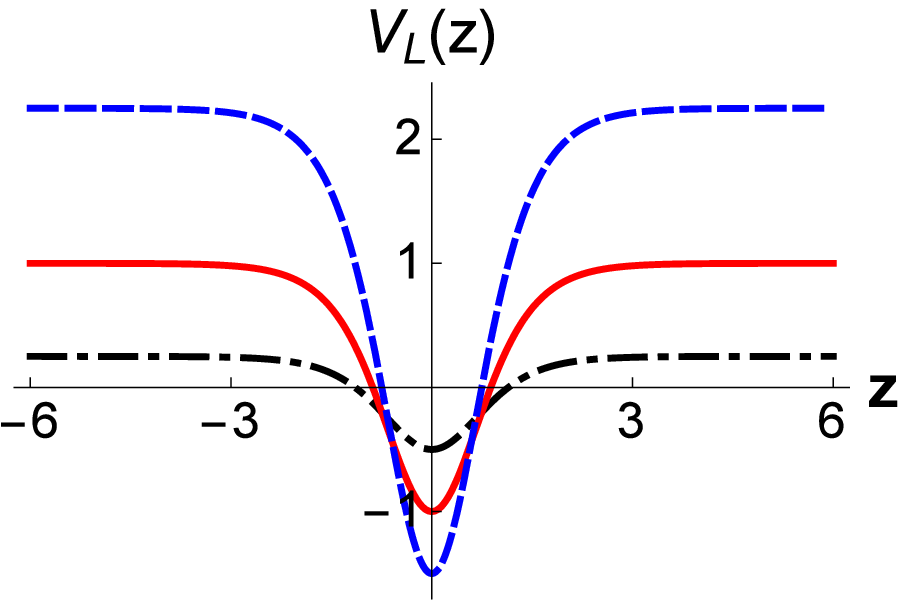}}
\subfigure[$c=1, q=1$]{
\includegraphics[width=5cm,height=3cm]{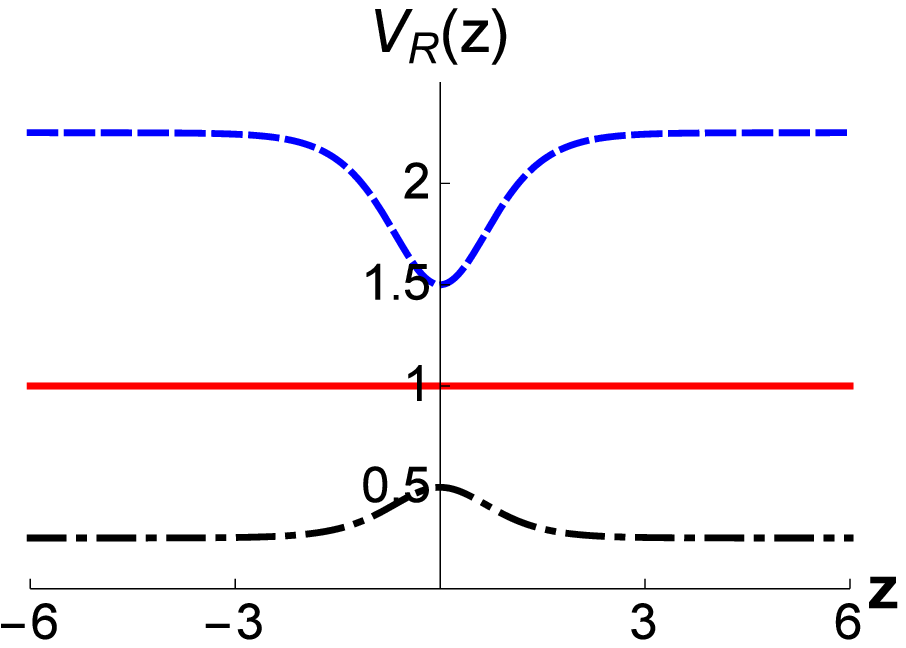}}\\
\subfigure[$c=1, q=2$]{
\includegraphics[width=5cm,height=3cm]{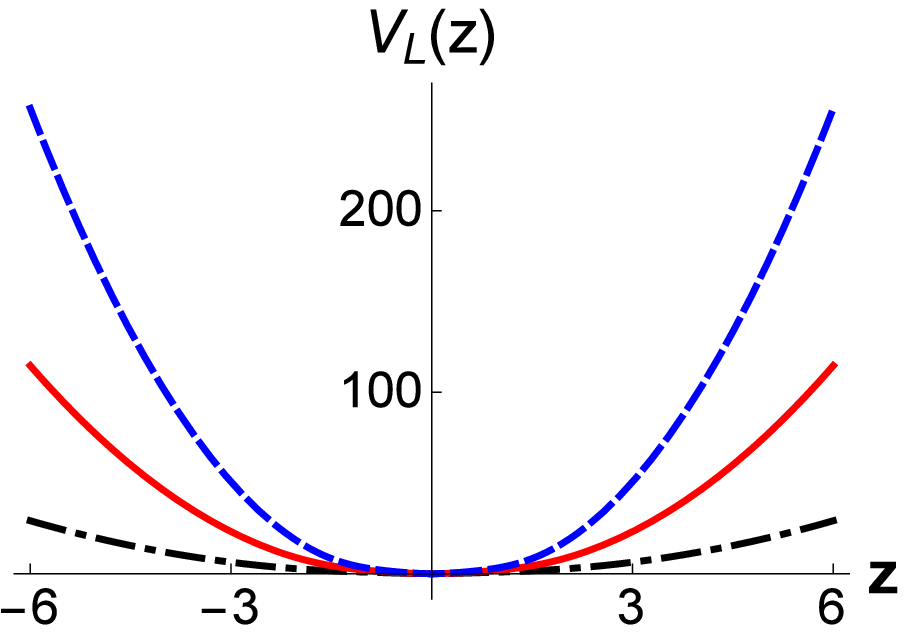}}
\subfigure[$c=1, q=2$]{
\includegraphics[width=5cm,height=3cm]{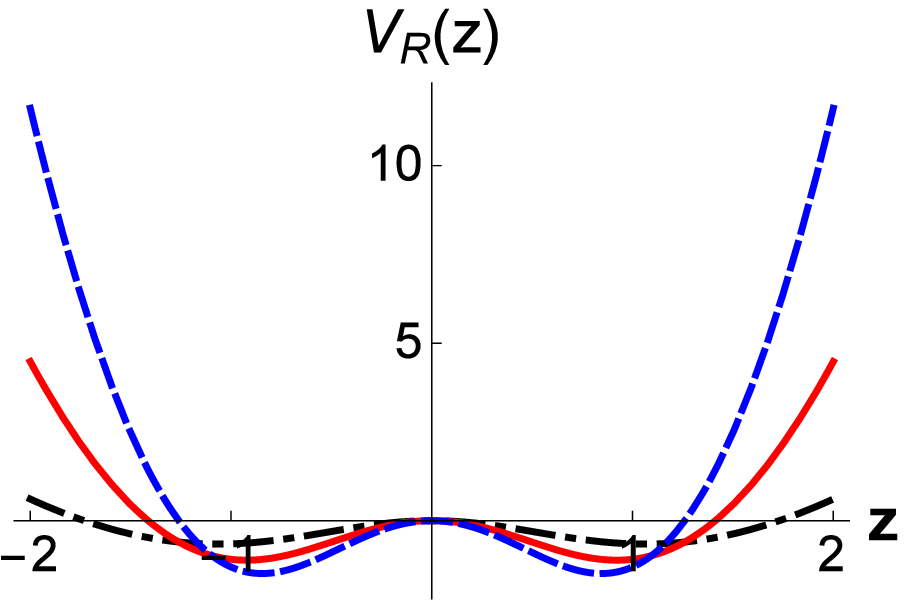}}\\
\subfigure[$c=1, q=3$]{
\includegraphics[width=5cm,height=3cm]{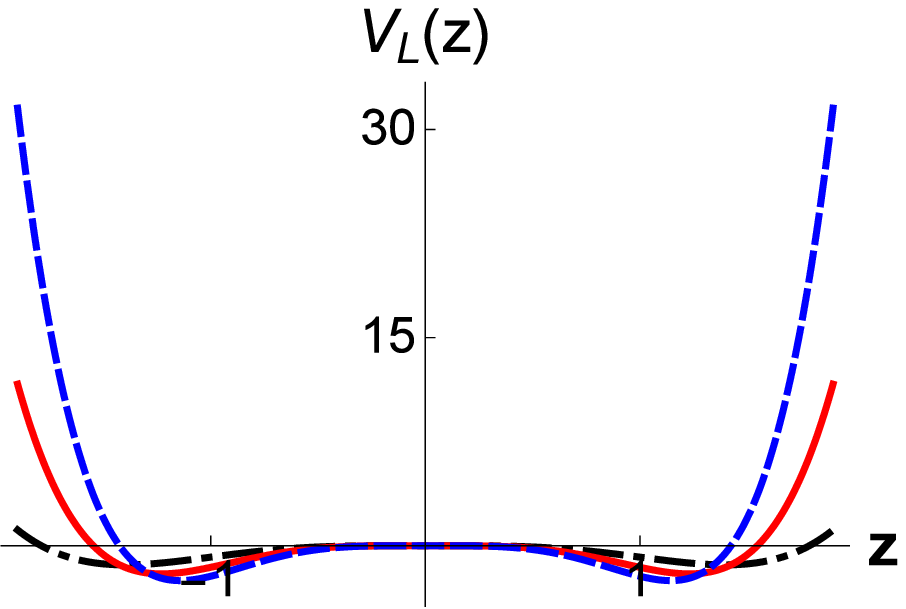}}
\subfigure[$c=1, q=3$]{
\includegraphics[width=5cm,height=3cm]{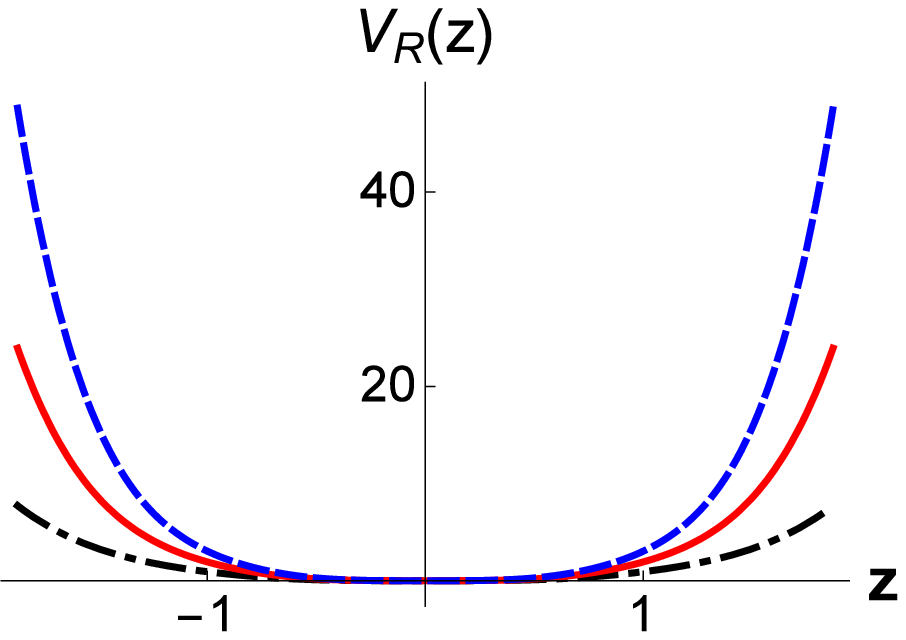}}\\
\end{center}
\caption{
Plots of the effective potentials in (\ref{Vtype2}) with the coupling $F(R)=\ln ^q\left(\text{sech}(\frac{2 \sqrt{8 c^2-R}}{ \sqrt{20 c^2+R}} )\right)$ for $c=1$, $q = 1,~2$  and  different values of $\eta$. The parameters are set to $\eta=0.5$ (black dot dashed line), 1.0 (red line), 1.5 (blue dashed line).
}
\label{VLVR2-2}
\end{figure}

The coupling parameter $\eta$ can affect the shapes of the effective potentials, especially for the case of $q=1$, which can be seen in Fig.~\ref{VLVR2-2}. For $q=1$ and $\eta=1$, the effective potential $V_R$ is flat. For $q=1$ and $\eta > (<) 1$, $V_R$ has a potential well (barrier). For $q\ge 2$, the slopes of the effective potentials $V_L$ and $V_R$ will increase with the coupling parameter $\eta$.

We have the zero modes as follows:
  \begin{eqnarray}
    f_{L0,R0}\propto \exp\left[\pm\eta\ln^q\left(\text{sech}(cz)\right)\right].
  \end{eqnarray}
In order to localize the zero mode on the brane, the following normalization condition should be satisfied
  \begin{eqnarray}
    \int_{0}^\infty \exp\left[\pm2\eta\ln^q\left(\text{sech}(cz)\right)\right]dz<\infty.
    \label{normalizationcondition2}
  \end{eqnarray}
It is easy to see that for $q=1$, the left-chiral zero mode can be localized on the brane, while the right-chiral zero mode cannot. For $q\geq2$, with the transformation $e^{cz}=t$, the integral in the formula (\ref{normalizationcondition2}) becomes  {$\int_1^{\infty}\frac{1}{c\,t}(\frac{2}{t+{1}/{t}})^{\pm2\eta\ln^{q-1}(\frac{2}{t+{1}/{t}} )}dt$}, whose integrand approaches $\frac{1}{c\,t}(\frac{2}{t})^{\pm2\eta(-\infty)^{q-1}}$ when $t\rightarrow\infty$. So whether the zero mode can be localized on the brane depends on $q$: the left-chiral (right-chiral) fermion zero mode can be localized on the brane if $q$ is odd (even). This is coincident with the analysis of the effective potentials.

\subsubsection{$F(R)={-\left(\frac{4(8c^2-R)}{20c^2+R}\right)^q }$}

At last we take $F(R)={-\left(\frac{4(8c^2-R)}{20c^2+R}\right)^q }$. With the same procedure, we can also get the effective potentials as follows:
  \begin{eqnarray}
    V_{L,R}(z)= {2  q\eta c^2  \left[2 q \eta  (cz)^{2 q}\mp(2q-1)\right] (cz)^{2 q-2}  },
    \label{Vtype3}
  \end{eqnarray}
for which one has
  \begin{eqnarray}
    V_{L,R}(0)&=& \left\{
                \begin{aligned}
                 \mp 2 c^2\eta,~ q=1 \\
                  0, ~q\geq2\\
                \end{aligned}
                \right.
                \nonumber \\
    V_{L,R}(\pm \infty)&=& \infty.
  \end{eqnarray}
\begin{figure}[htb]
\begin{center}
\subfigure[$c=1, q=1, \eta=0.2$]{
\includegraphics[width=5cm,height=3cm]{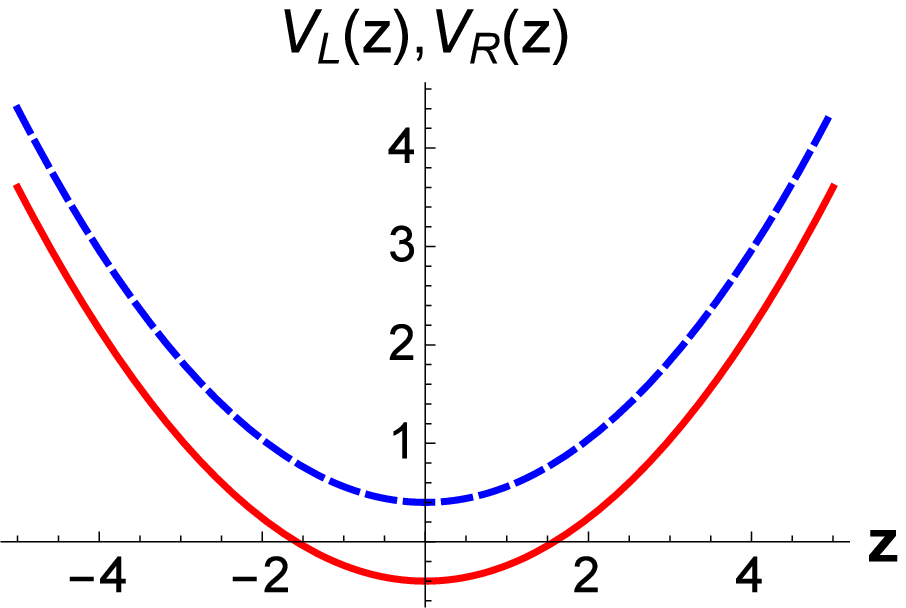}}
\subfigure[$c=1, q=2, \eta=0.2$]{
\includegraphics[width=5cm,height=3cm]{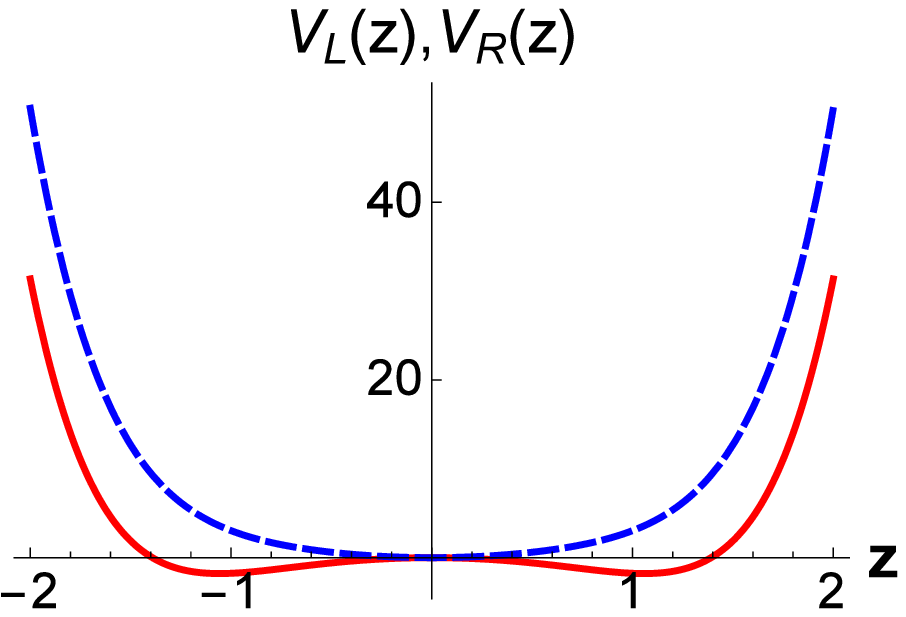}}
\subfigure[$c=1, q=3, \eta=0.02$]{
\includegraphics[width=5cm,height=3cm]{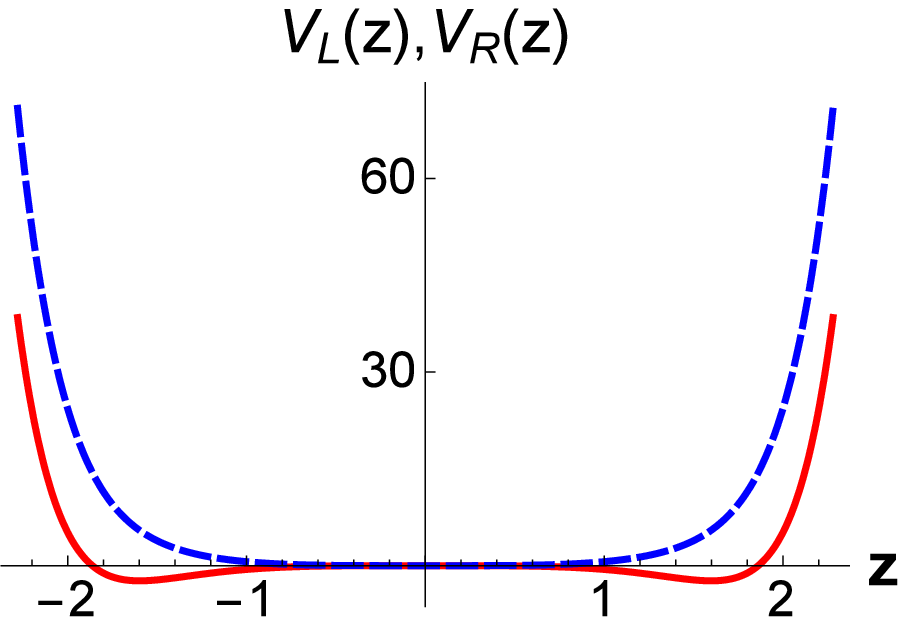}}
\end{center}
\caption{
Plots of the effective potentials  $V_{L}(z)$ (red curves) and $V_{R}(z)$ (blue dashed curves) in (\ref{Vtype3}) with the coupling $F(R)={-\left(\frac{4(8c^2-R)}{20c^2+R}\right)^q }$ for $c=1$ and different values of $q$ and $\eta$.
}
\label{VLVR3-1}
\end{figure}

\begin{figure}[htb]
\begin{center}
\subfigure[$c=1, q=1$]{
\includegraphics[width=5cm,height=3cm]{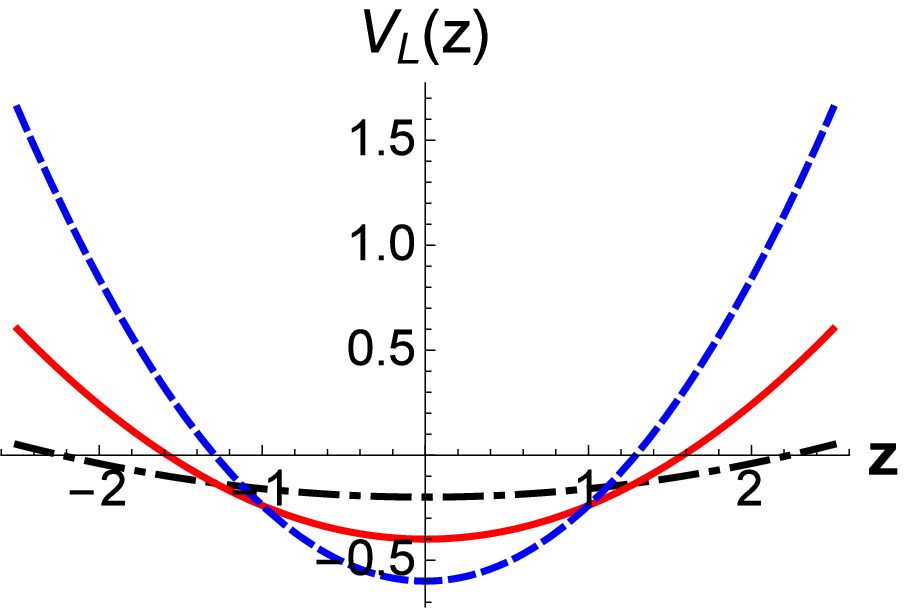}}
\subfigure[$c=1, q=1$]{
\includegraphics[width=5cm,height=3cm]{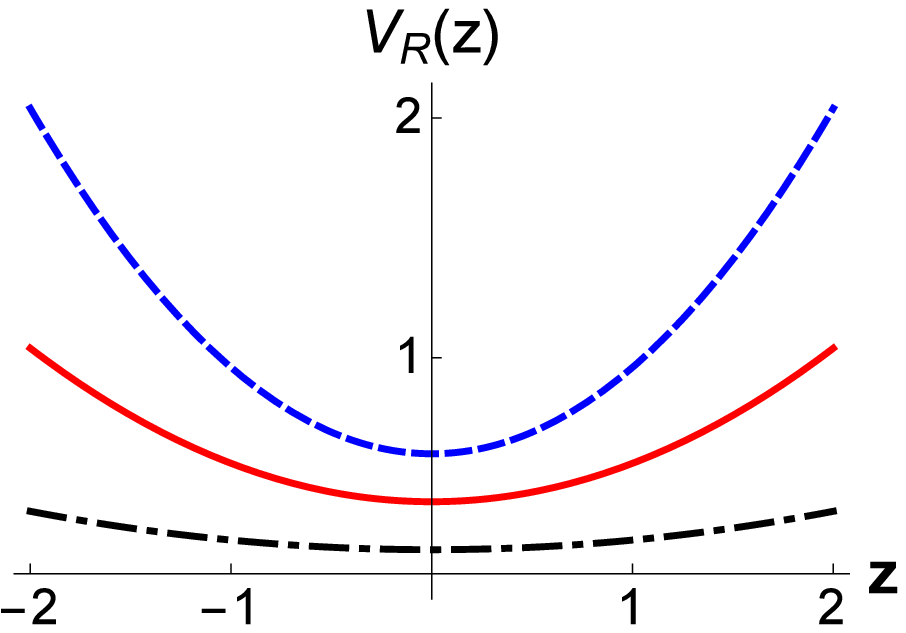}}\\
\subfigure[$c=1, q=2$]{
\includegraphics[width=5cm,height=3cm]{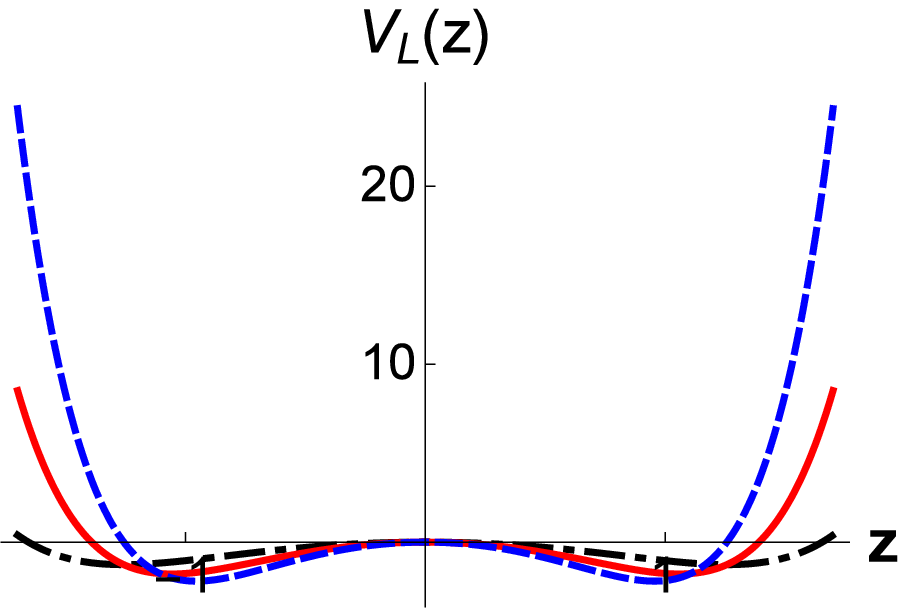}}
\subfigure[$c=1, q=2$]{
\includegraphics[width=5cm,height=3cm]{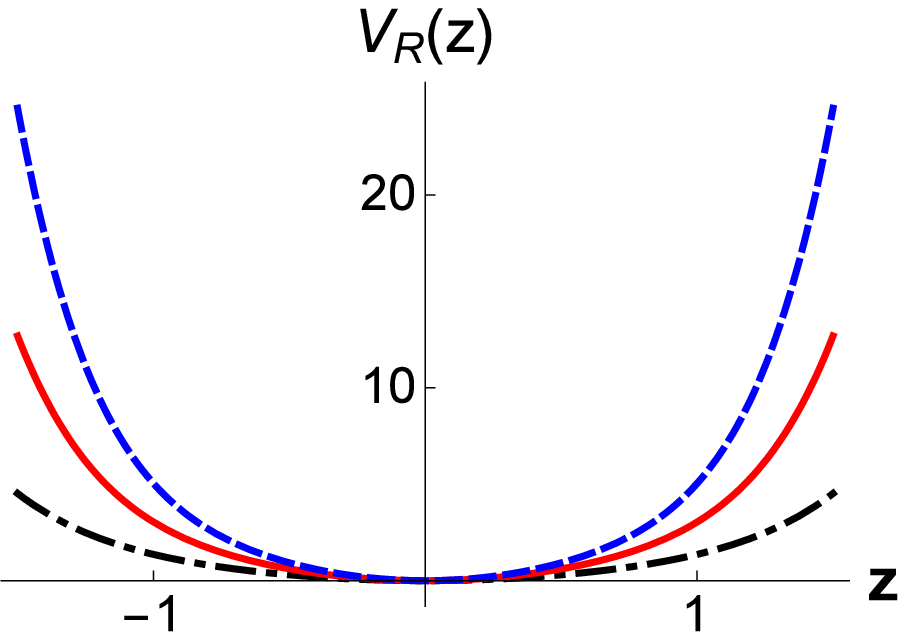}}
\end{center}
\caption{
Plots of the effective potentials in (\ref{Vtype3}) with the coupling $F(R)={-\left(\frac{4(8c^2-R)}{20c^2+R}\right)^q }$ for $c=1$ and $q = 1,~2$  and different values of $\eta$. The parameters are set to $\eta=0.1$ (black dot dashed line), 0.2 (red line), 0.3 (blue dashed line).
}
\label{VLVR3-2}
\end{figure}

\begin{figure}[htb]
\begin{center}
\subfigure[~$m^2_{L_n}$]{
\includegraphics[width=6cm,height=3.6cm]{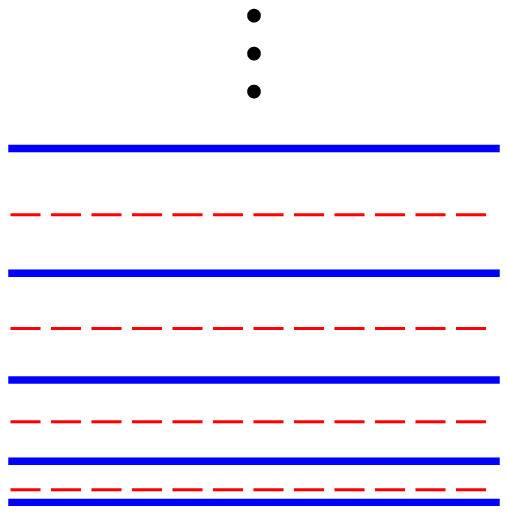}}
\subfigure[~$m^2_{R_n}$]{
\includegraphics[width=6cm,height=3.6cm]{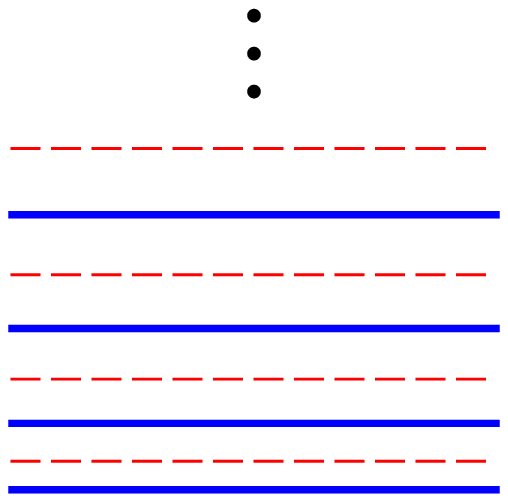}}
\end{center}
\caption{
The mass spectra $m^2_{L_n,R_n}$ of the left- and right-chiral fermions with  even (blue curves) and odd  parity (red dashed curves) in (\ref{Vtype3}) with the coupling $F(R)={-\left(\frac{4(8c^2-R)}{20c^2+R}\right)^q }$ for $c=1, q=2$ and $\eta=0.3$.
}
\label{mass3}
\end{figure}

Plots of the effective potentials are shown in Figs.~\ref{VLVR3-1} and ~\ref{VLVR3-2}. We also set the coupling constant $\eta$ to be a positive number. For $q=1$, we can get $V_L(0)<0$ and $V_R(0)>0$, and the values of the effective potentials diverge when $z\rightarrow\infty$. Therefore, both the left- and right-chiral fermion KK modes have the discrete mass spectra. We get the analytical mass spectra of the left- and right-chiral fermion KK modes in the case of $q=1$ and $c=1$:
\begin{eqnarray}
    m^2_{L_n}&=& 4n\eta, ~(n=0,1,2,3,\cdots), \\
    m^2_{R_n}&=& 4n\eta, ~(n=~~~1,2,3,\cdots).
\end{eqnarray}
As can be seen from the above equation, when $q$ takes 1, there are infinite KK bound states, only the left-chiral zero mode can be localized on the brane and both the mass spectra $m^2_{L_n(R_n)}$ intervals are the constant $4\eta$.

When $q\geq2$, $V_{L,R}(0)=0$ and both the effective potentials will diverge at the boundaries of the conformal coordinate, so there is a discrete mass spectrum. The left-chiral effective potential $V_L(z)$ has a double well, which could lead to the localization of the zero mode. However, for the right-chiral effective potential $V_R(z)$ with one well, the zero mode cannot be localized on the brane.
The mass spectra of lower KK modes $m^2_{L_n,R_n}$ for $q=2,\eta=0.3$ are shown in Fig.~\ref{mass3} and are listed as follows:
\begin{eqnarray}
 m_{L_n}^2 &=& (0, 2,6.5, 12.57,19.375,27.5, 36.25, 45.5,56,\cdots),
     \\
     m_{R_n}^2 &=& (~~~2,6.5, 12.57,19.375,27.5, 36.25, 45.5,56,\cdots).
\end{eqnarray}

The left- and right-chiral zero modes are
  \begin{eqnarray}
    f_{L0,R0}\propto \exp{  \left(\mp \eta (cz)^{2 n}\right)}.
  \end{eqnarray}
The corresponding normalization condition is given by
  \begin{eqnarray}
    \int^\infty_{-\infty}\exp{  \left(\mp 2\eta (cz)^{2 n}\right)}dz< \infty.
    \label{normalizationcondition3}
  \end{eqnarray}
From the above formula (\ref{normalizationcondition3}), we can easily realize that for positive $\eta$ and any positive integer $n$, the left-chiral zero mode satisfies the normalization condition and therefore can be localized on the brane, while the right-chiral one cannot.\\

\subsection{Brane model II}

In this subsection, we consider another kind of brane generated by a single scalar field. We assume that the warp factor in the conformal coordinate is $A(z)=\ln(\text{sech}(kz))$.
From the coordinate transformation  (\ref{coordinateTransformation}), the relation between the  physical coordinate and the conformal coordinate reads
\begin{eqnarray}
  y=\frac{2}{k} \text{arctan}\left(\text{tanh}\left(\frac{kz}{2}\right)\right).
\end{eqnarray}
Therefore, the warp factor in the physical coordinate is given by
\begin{eqnarray}
  A(y)=\ln\big(\cos (ky)\big),
\end{eqnarray}
which indicates that there is a horizon at $y=\pm \pi/(2k)$ or $z=\pm \infty$.
Through the field equations (\ref{fieldequation}), the scalar field and the corresponding potential are found to be
\begin{eqnarray}
  \phi (y) &=& \sqrt{\frac{3}{2}} \ln \left(\tan \left(\frac{1}{2}  k y+\frac{1}{4}\pi \right)\right),\label{phi2}\\
  V(\phi) &=& \frac{3}{8} k^2 \left(5-3 \cosh \left(2 \sqrt{\frac{2}{3}} \phi \right)\right).\label{v2}
\end{eqnarray}
Therefore, there exists a brane solution generated by a real scalar field with the warp factor $A(z)=\ln(\text{sech}(kz))$. The scalar field $\phi(z)$ in the conformal coordinate can also be given with the coordinate transformation (\ref{coordinateTransformation}),
\begin{eqnarray}
  \phi(z)=\sqrt{\frac{3}{2}} kz.
\end{eqnarray}

\begin{figure}[htb]
\begin{center}
\subfigure[~$m^2_{L_n}$]{
\includegraphics[width=6cm,height=3.6cm]{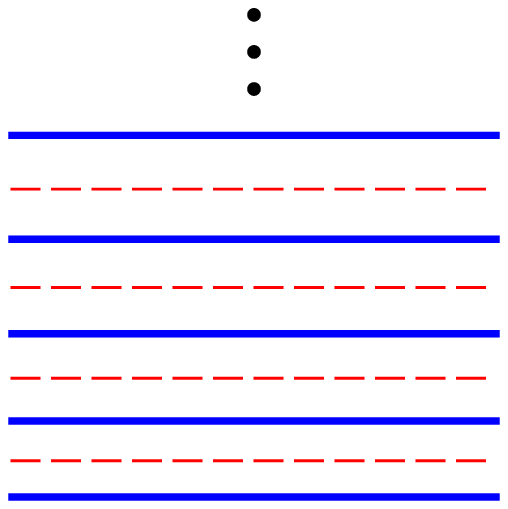}}
\subfigure[~$m^2_{R_n}$]{
\includegraphics[width=6cm,height=3.6cm]{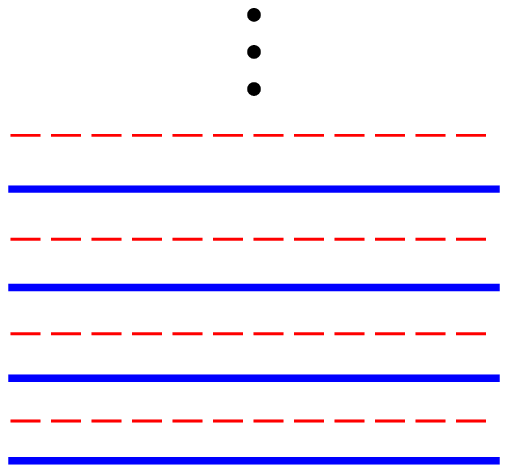}}
\end{center}
\caption{
The mass spectra $m^2_{L_n,R_n}$ of the left- and right-chiral fermions with even (blue curves) and odd  parity (red dashed curves) in (\ref{Vtype4}) with the coupling $F(R)=R$ for $c=1, n=1$ and $\eta=1$.
}
\label{mass4}
\end{figure}

In this case, we get $R=2 k^2 [7-3 \cosh (2 k z)]$ and take $F(R)=R$. Then the corresponding effective potentials are
\begin{eqnarray}
  V_{L,R}(z)&=&24 \eta  k^4 \left[6 \eta  k^2 \sinh ^2(2 k z)\mp\cosh (2 k z)\right].
 \label{Vtype4}
\end{eqnarray}
The values of the left- and right-chiral effective potentials at $z=0$ and $z\rightarrow \pm\infty$ are
\begin{eqnarray}
    V_{L,R}(0)= \mp 24k^4\eta,~~
    V_{L,R}(\pm \infty)= +\infty.
\end{eqnarray}
We only consider positive coupling constant $\eta$. In this condition, $V_L(0)<0$ and $V_R(0)>0$, and both the left- and right-chiral potentials approach infinity when $z \rightarrow \pm\infty $. Therefore, for the left-chiral fermion, the zero mode could be localized on the brane, and there is a discrete mass spectrum. While for the right-chiral fermion, the zero mode cannot be localized on the brane and there is a series of masssive bound KK modes with $m_n^2>24k^4\eta$. The mass spectra $m^2_{L_n,R_n}$  are shown in Fig. \ref{mass4} and listed as follows:
\begin{eqnarray}
 m_{L_n}^2 &=& (0, 2,48.75,102.5,160, 220,282.5, 347.5,415, 635,\cdots),\\
     m_{R_n}^2 &=& (~~~2,48.75,102.5,160, 220,282.5, 347.5,415, 635,\cdots),
\end{eqnarray}
for $\eta=1$ and $q=2$, which shows that there are a localized left-chiral fermion zero mode with even parity and an infinite number of KK bound states of left- and right-chiral fermions in pairs with the same mass but opposite parity.

\begin{figure}[htb]
\begin{center}
\subfigure[$k=1,\eta=1$]{
\includegraphics[width=5cm,height=3cm]{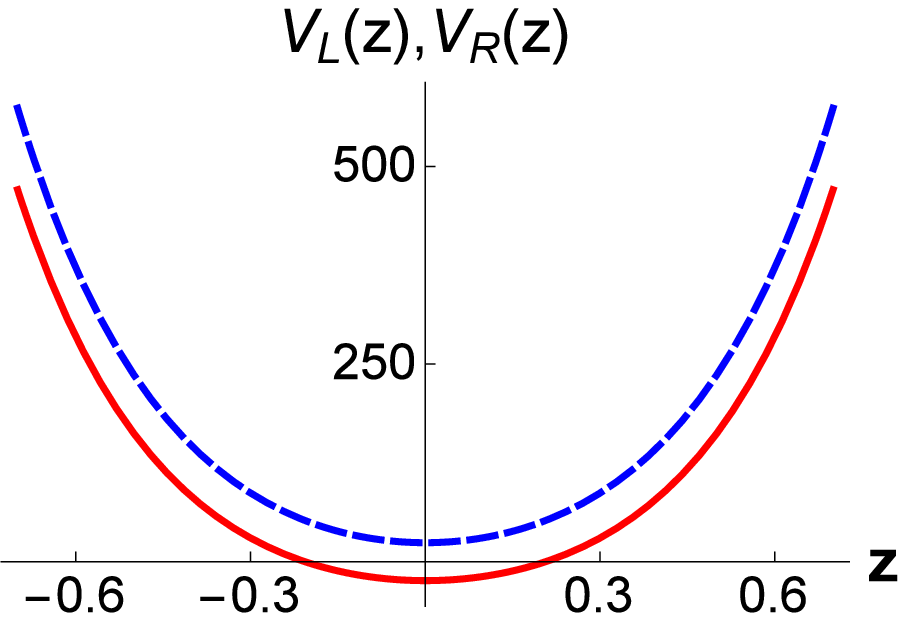}}
\subfigure[$k=1,\eta=2$]{
\includegraphics[width=5cm,height=3cm]{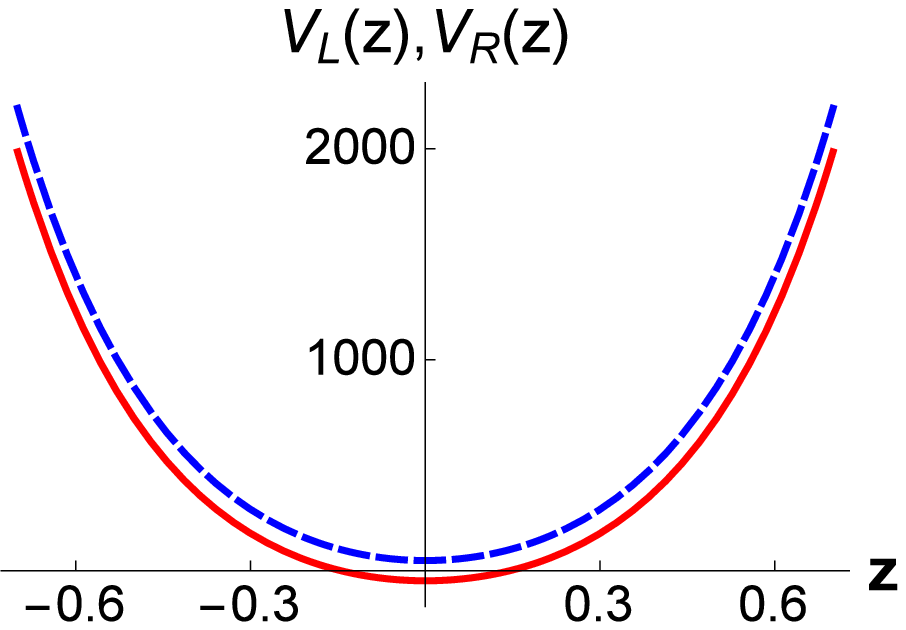}}
\subfigure[$k=2,\eta=1$]{
\includegraphics[width=5cm,height=3cm]{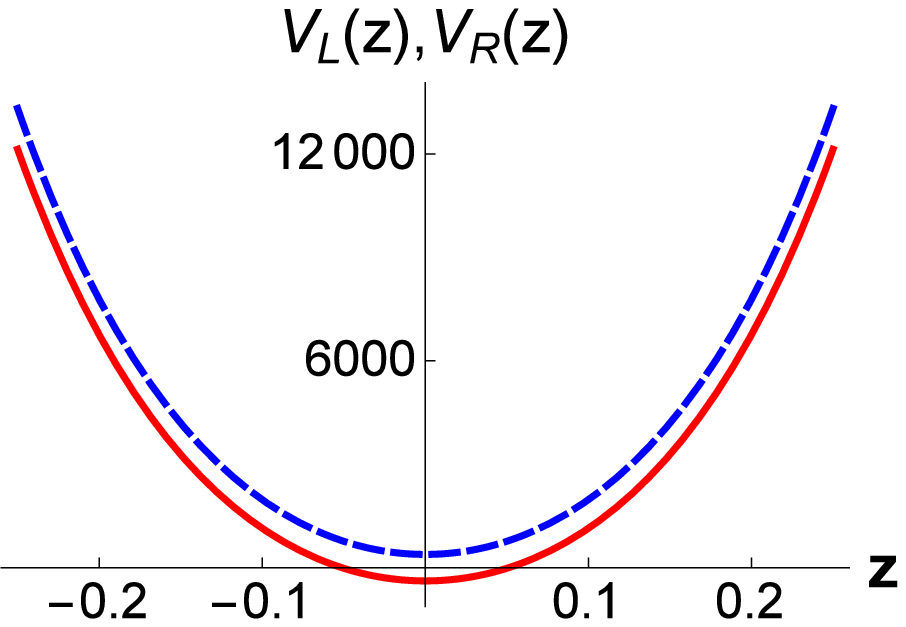}}
\end{center}
\caption{
Plots of the effective potentials $V_{L}(z)$ (red curves) and $V_{R}(z)$ (blue dashed curves) in (\ref{Vtype4}) with the coupling $F(R)=R$ for different values of $k$ and $\eta$.
}

\label{VLVR4}
\end{figure}

The left- and right-chiral zero modes are
\begin{eqnarray}
    f_{L_0,R_0}(z)\propto e^{\pm \eta R}=\exp\left[\pm 2\eta k^2 (7-3 \cosh (2 k z))\right].
\end{eqnarray}
We can easily find that only the left-chiral zero mode satisfies the normalization condition.

\subsection{Brane model III}

In this part, we mainly focus on a five-dimensional thick brane model in pure metric $f(R)$ gravity without scalar fields. On account of the absence of the scalar fields, it is necessary to introduce the new coupling $\eta\bar{\Psi}\Gamma^M\partial_M{F(R)}\gamma^5\Psi$ in order to localize a fermion.

In this example, we consider a simple thick brane model of the pure geometry in the five-dimensional spacetime.
The action of the this system is \cite{Zhong2016}
\begin{eqnarray}
\label{pureaction}
S=\frac{1}{2\kappa_5^2}\int d^5x\sqrt {-g}f(R),
\end{eqnarray}
where $\kappa_5^2=8\pi G^{(5)}$ is the five-dimensional gravitational coupling constant. The metric still takes the previous form (\ref{Metric}). One thick brane solution \cite{Zhong2016} is
\begin{eqnarray}
A(y)&=& -\ln (\cosh (w)) \\
f(R)&=&\frac{4}{7} \left(6 k^2+R\right)\cosh (\alpha(w(R)) )
-\frac{2}{7} k^2 \sqrt{480-\frac{36 R}{k^2}-\frac{3 R^2}{k^4}} \sinh (\alpha(w(R)) ),
\end{eqnarray}
where $w=ky$, $w(R)=\pm\textrm{arcsech}\left(\frac{\sqrt{20 +R/k^2}}{2\sqrt{7}}\right)$ and $\alpha (w)\equiv 2 \sqrt{3} \arctan \left(\tanh \left(\frac{w}{2} \right)\right)$.

Compared with the previous brane model I, one can easily find that the two metrics are very similar except the difference of a rescaling factor. Therefore, we can also take different forms of $F(R)$ to obtain three similar  effective potentials as in the model I. More specifically, we can take $F(R)=\ln ^q\left(\frac{20 k^2+R}{28 k^2}\right)$, $F(R)=\ln ^q\left(\text{sech}(\frac{\sqrt{8 k^2-R}}{\sqrt{20 k^2+R}})\right)$, and $F(R)=-\left(\frac{8 k^2 - R}{20 k^2 + R}\right)^ q$, which correspond to the three forms of $F(R)$ in the first example, to localize the fermion zero mode, respectively. In particular, when $q=1${\color{green},} one can get the corresponding volcano, PT-like and harmonic-like effective potentials. Besides, we can also get the similar mass spectra. Therefore, fermions can be localized on the branes even without the appearance of background scalar fields with our new coupling mechanism.

\section{Discussion and Conclusion}\label{Section4}
In this paper, we proposed a new localization mechanism in order to localize fermions on branes. In the past, in order to localize the zero mode of a fermion on a brane generated by a background scalar field $\phi$, the coupling between the fermion and the background scalar field was introduced. A usual choice is the Yukawa coupling  $\eta \phi\bar {\Psi}\Psi  $.
 However, in this mechanism, the scalar field must be an odd function of the extra dimension. Recently, the authors of Ref.~\cite{Liu2014} have presented another localization mechanism with the coupling $\eta\bar{\Psi}\Gamma^M\partial_M{F(\phi)}\gamma^5\Psi$, which can be applied   to the case of an even or odd scalar field. Inspired by the coupling mechanism and considering the fact that the scalar curvature is an even function of the extra dimension because of $Z_2$ symmetry, we presented a new coupling between the  spacetime curvature and the fermion, $\eta\bar{\Psi}\Gamma^M\partial_M{F(R)}\gamma^5\Psi$, to localize the fermion on the brane. More importantly, this new coupling is necessary for a brane model without background scalar fields. In such brane models, because of the lack of scalar fields, the two coupling mechanisms mentioned earlier are no longer applicable. However, our newly proposed coupling mechanism can work very well.
Then, we explored the new coupling mechanism to study the localization of fermions and obtain the corresponding mass spectra  through three specific brane models. We only considered the case of $\eta>0$ in this paper.

In the first brane model, the brane is generated by a single kink scalar field. For this model, we mainly investigated the localization and mass spectra of the fermion by considering three different forms of $F(R)$.
Firstly, we considered the case of $F(R)=\ln ^q\left(\frac{20 c^2+R}{52 c^2-3R}\right)$, for which the effective potentials have a volcano shape and so there is no bound KK mode except the zero modes.  For the case of $q=1$, only the zero mode of the left-chiral fermion can be localized on the brane under the condition $\eta>1/4$. For any positive odd (even) $q\geq2$, there appears a double well near $z=0$ for the effective potential of the left-chiral (right-chiral) KK modes and only the left-chiral (right-chiral) zero mode can be localized on the brane.

Secondly, we took $F(R)=\ln ^q\left(\text{sech}(\frac{2 \sqrt{8 c^2-R}}{ \sqrt{20 c^2+R}} )\right)$. For $q=1$, the effective potentials are a positive constant at infinity, and so they are PT-like potentials. The left-chiral fermion zero mode can be localized on the brane because of the negative values of $V_{L}(z)$ around $z=0$. In addition to the left-chiral fermion zero mode, there may exist the left- and right-chiral bound KK modes with the same mass but opposite parity when $\eta>1$.
Furthermore, for $q\geq 2$, the effective potentials diverge at infinity. Therefore, there are discrete mass spectra for both the left- and right-chiral fermions. Besides, when $q$ is an odd (even) number greater than  1, the left-chiral (right-chiral) fermion zero mode is localized on the brane.

Thirdly, we chose $F(R)={-\left(\frac{4(8c^2-R)}{20c^2+R}\right)^q }$. For the case of $q=1$, the effective potentials are the harmonic-like ones which generate the equal interval mass spectrum and the left-chiral fermion zero mode can be localized on the brane. When $q\geq2$, there is a discrete mass spectrum, and the left-chiral fermion zero mode is also localized on the brane.

In the second brane model, we considered another kind of brane generated by a single scalar field. In this example,  we explored a special case, that is, $F(R)=R$, with which we can also obtain the localized left-chiral fermion zero mode and an infinite number of bound KK modes of the left- and right-chiral fermions with the same mass but opposite parity.

At last, we mainly investigated a pure geometric thick brane model.  As the first example, we briefly discussed three forms of $F(R)$ to localize the fermion zero mode. These results show that the problem of fermion localization on the pure geometric thick brane can be solved successfully with the introduction of this new coupling mechanism.

\section*{Acknowledgement}

This work was supported by the National Natural Science Foundation of China (Grants Nos. 11522541, 11375075, and 11205074), and the Fundamental Research Funds for the Central Universities (Grants No. lzujbky-2016-k04).


%


\end{document}